\documentclass[amssymb,amsfonts,aps,prb,twocolumn]{revtex4-1}

\usepackage{graphicx}
\usepackage{dcolumn}
\usepackage{bm}
\usepackage[normalem]{ulem}
\usepackage{xcolor}
\usepackage{comment}
\usepackage{amsmath}
\usepackage{siunitx}
\usepackage{dsfont}
\usepackage{float}
\usepackage{placeins}

\newcommand{\up}{\uparrow}
\newcommand{\dn}{\downarrow}
\DeclareSIUnit\angstrom{\text{Å}}

\begin{document}

\title{
Broken-symmetry magnetic phases in two-dimensional triangulene crystals
}

\author{
G. Catarina$^{1,2}$, J. C. G. Henriques$^{1,3}$, A. Molina-S\'anchez$^{4}$, A. T. Costa$^1$, J. Fern\'andez-Rossier$^{1,}$
}
\altaffiliation{
On permanent leave from Departamento de F\'isica Aplicada, Universidad de Alicante, 03690 San Vicente del Raspeig, Spain.
}

\affiliation{
$^1$International Iberian Nanotechnology Laboratory (INL), Av. Mestre Jos\'e Veiga, 4715-330 Braga, Portugal
}
\affiliation{
$^2$Current address: nanotech@surfaces Laboratory, Empa---Swiss Federal Laboratories for Materials Science and Technology, 8600 D\"{u}bendorf, Switzerland
}
\affiliation{
$^3$Universidade de Santiago de Compostela, 15782 Santiago de Compostela, Spain
}
\affiliation{
$^4$Institute of Materials Science (ICMUV), University of Valencia, Catedr\'{a}tico Beltr\'{a}n 2, E-46980 Valencia, Spain
}

\date{\today}

%------------------------------------------------------------------------
\begin{abstract} 
We provide a comprehensive theory of magnetic phases in two-dimensional triangulene crystals, using both Hubbard model and density functional theory (DFT) calculations.   
We consider centrosymmetric and non-centrosymmetric triangulene crystals. 
In all cases, DFT and mean-field Hubbard model predict the emergence of broken-symmetry antiferromagnetic (ferrimagnetic) phases for the centrosymmetric (non-centrosymmetric) crystals.   
This includes the special case of the [4,4]triangulene crystal, whose non-interacting energy bands feature a gap with flat valence and conduction bands.
We show how the lack of contrast between the local density of states of these bands, recently measured via scanning tunneling spectroscopy, is a natural consequence of a broken-symmetry N\'eel state that blocks intermolecular hybridization.   
Using random phase approximation, we also compute the spin wave spectrum of these crystals, including the recently synthesized [4,4]triangulene crystal.
The results are in excellent agreement with the predictions of a Heisenberg spin model derived from multi-configuration calculations for the unit cell. 
We conclude that experimental results are compatible with an antiferromagnetically ordered phase where each triangulene retains the spin predicted for the isolated species.
\end{abstract}

\maketitle

%------------------------------------------------------------------------
\section{Introduction}
\begin{figure*}
    \centering
    \includegraphics[width=0.75\linewidth]{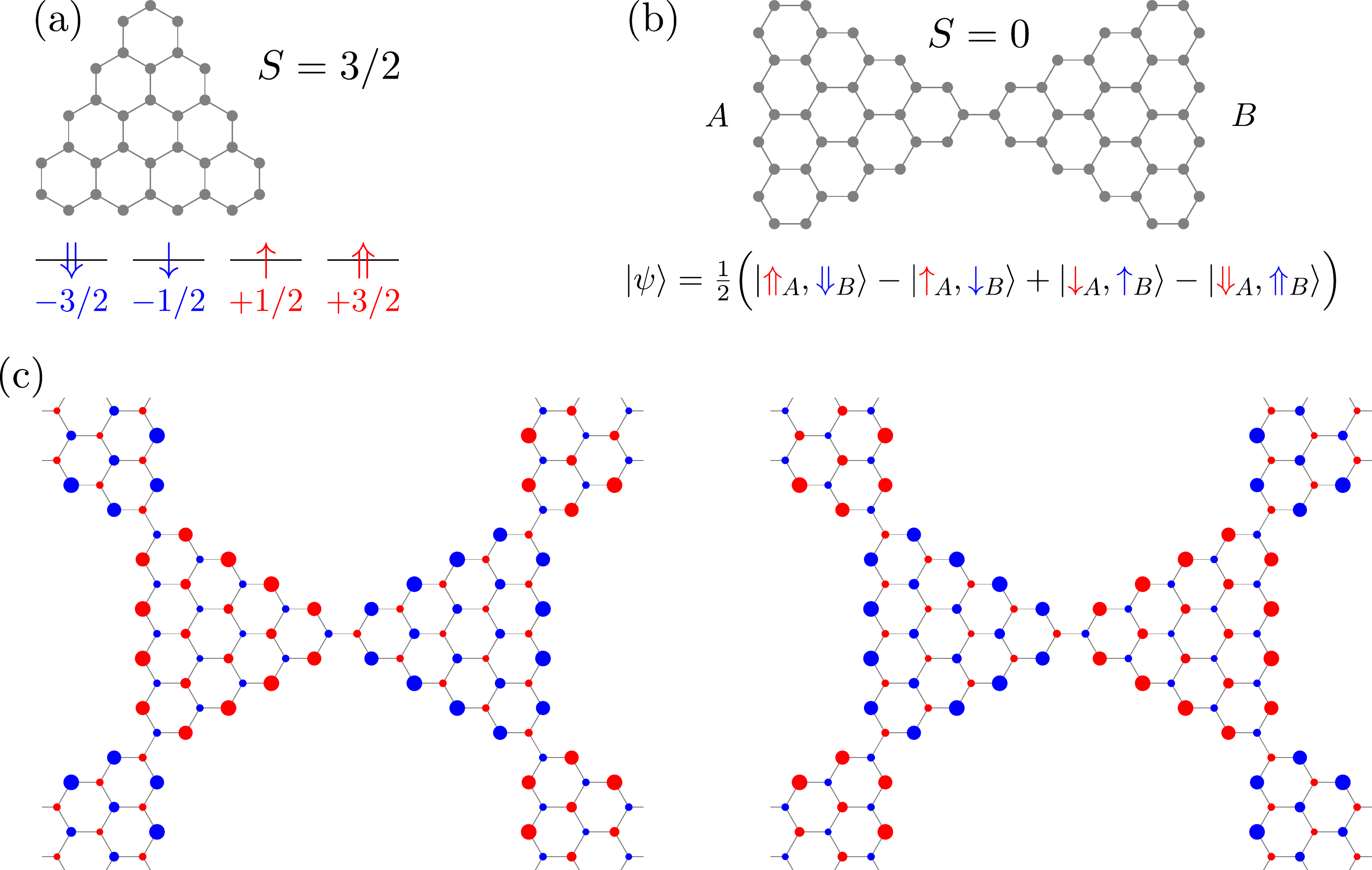}
    \caption{ 
    (a) [4]triangulene has a fourfold degenerate ground state with total spin $S=3/2$. 
    (b) [4]triangulene dimer is an open-shell singlet with an entangled wave function as a result of antiferromagnetic intermolecular coupling. 
    (c) Two examples of broken-symmetry N\'eel states for [4,4]triangulene 2D crystals, obtained with a collinear mean-field Hubbard model. 
    The size of the circles represents the magnitude of the local moments, with red/blue colors denoting spin-$\up$/$\dn$.
    Total spin is no longer a good quantum number. 
    }
    \label{fig:1}
\end{figure*}

Triangulenes are graphene fragments with the shape of an equilateral triangle, terminated with zigzag edges and of various sizes, customarily defined in terms of the number $n$ of benzenes in a given edge\cite{clar53,fernandez07,su2020}.  
According to single-particle theory, $[n]$triangulenes host $n-1$ non-bonding half-filled zero modes\cite{fernandez07}. 
Coulomb interactions favor the maximal spin configuration, very much like the Hund's first rule in atoms, so that $[n]$triangulenes are predicted\cite{fernandez07,wang2008,wang09,yazyev10,ortiz19} to have a ground state with total spin $S=\frac{n-1}{2}$ (see Fig.~\ref{fig:1}a), consistent with Lieb's theorem for the Hubbard model for bipartite lattices at half-filling\cite{Lieb1989}, and in agreement with Ovchinnikov's rule\cite{ovchinnikov78}.

The highly reactive nature of radicals hampered the experimental study of triangulenes for several decades. 
This situation has radically changed with the advent of on-surface synthesis\cite{cai2010,Song2021} and experimentation in ultra-high vacuum.  
Therefore, triangulenes of various sizes ($n=2,3,4,5,7$) have been synthesized, both in isolated form\cite{pavlivcek2017,su2019,mishra2019b,mishra2021b,turco23}, and also forming dimers\cite{mishra2020}, rings\cite{Mishra2021,hieulle2021}, chains\cite{Mishra2021}, and, very recently, small-size two-dimensional (2D) lattices\cite{delgado23}. 

Using inelastic electron tunneling spectroscopy\cite{Ortiz2020}, zero-bias Kondo resonances in individual $[2]$triangulenes\cite{turco23}, as well as spin excitations in $[3]$triangulene dimers\cite{mishra2020}, rings\cite{Mishra2021,hieulle2021} and chains with more than 40 units\cite{Mishra2021} have been observed. 
These experiments provide strong evidence that these zero- and one-dimensional supramolecular structures remain open-shell and their low-energy electronic properties can be accounted for by spin Hamiltonians with antiferromagnetic interactions (Fig.~\ref{fig:1}b). 

Spin-restricted density functional theory (DFT) calculations of $[n,m]$triangulene crystals---i.e., honeycomb 2D crystals whose unit cell is made of a pair of triangulenes with sizes $n$ and $m$---show the formation of $n+m-2$ weakly dispersive energy bands\cite{ortiz22}. 
Using tight-binding models, it has been shown\cite{ortiz22} that these bands are made of linear combinations of the in-gap zero modes of the triangulenes, hybridized via third-neighbor hopping. 
Intermolecular hybridization splits the zero modes into bonding-antibonding pairs, promoting non-magnetic closed-shell electronic configurations. 
Therefore, in contrast with the case of isolated triangulenes, interactions need to overcome intermolecular hybridization in order to promote open-shell states.   
This is expected to be harder in the case of the $[4,4]$triangulene crystal, for which both spin-restricted DFT\cite{sethi2021} and tight-binding calculations predict a narrow-gap insulator, unlike the $[2,2]$ and $[3,3]$ cases, that feature Dirac cones at the Fermi energy.  
The synthesis of a $[4,4]$triangulene 2D lattice has been recently reported\cite{delgado23}, putting this specific system under the spotlight. 

In this work, we undertake a systematic study of the electronic properties of triangulene 2D crystals, focusing on the magnetic properties of their ground states. 
To do so, we go beyond the spin-restricted framework in the case of DFT, and beyond non-interacting tight-binding models.  
For that matter, we take the natural next step, doing spin-unrestricted DFT calculations and adding a Hubbard term to the tight-binding model used in previous work. 
The Hubbard model is treated at three levels of approximation: collinear mean-field theory, random phase approximation (RPA) and exact diagonalization of small structures in a restricted space of configurations, the so-called complete active space (CAS) method. 

Previous spin-unrestricted DFT calculations predict that [2,2]- (ref.\cite{zhou20}) and [3,3]- (ref.\cite{ortiz22}) triangulene 2D crystals should display antiferromagnetic order, with the two sublattices being polarized in opposite directions.  
The [4,4]triangulene crystal is different from [2,2] and [3,3] as it features a small band-gap when calculated both with spin-restricted DFT\cite{sethi2021,ortiz22,delgado23} and with the conventional single-orbital tight-binding model with third-neighbor hopping\cite{ortiz22}.  
On the basis of this narrow gap, an excitonic insulating phase has been proposed\cite{sethi2021}, taking as a reference-state the closed-shell non-magnetic ground state.  
A major goal of this manuscript is to address whether the [4,4]triangulene crystal also features an antiferromagnetic phase (Fig.~\ref{fig:1}c), and how this affects the size of the gap and the putative excitonic insulator.

The rest of the paper is organized as follows.  
In section~\ref{methods} we review the different theoretical methods used in this work. 
In section~\ref{CAS} we present our results for triangulene dimers within the CAS approach for the Hubbard model.  
These calculations allow us to derive the effective spin interactions, in the form of polynomials of the Heisenberg coupling, and to estimate the magnitude of the intermolecular exchange couplings.  
In section~\ref{2D} we present the results of collinear mean-field Hubbard and spin-unrestricted DFT calculations.  
The results are very similar, validating the Hubbard model, and systematically predict broken-symmetry magnetic phases as the ground state of triangulene 2D crystals.
In section~\ref{RPA} we present our Hubbard model RPA calculations of the spin waves for the [2,3], [3,3] and [4,4] crystals, and compare them with those obtained from the spin models derived from the Hubbard model CAS calculations. 
In section~\ref{LDOS} we discuss how the lack of contrast in the local density of states (LDOS) of the conduction and valence bands can be used to identify the emergence of broken-symmetry states, providing an explanation to recent experimental scanning tunneling microscope (STM) spectroscopy results\cite{delgado23}.  
In section~\ref{discuss} we present the conclusions. 

%------------------------------------------------------------------------
\section{Methods\label{methods}}
In this section we provide a brief description of the theoretical methods used throughout the paper.   

\subsection{Hubbard model}
Following previous work\cite{fujita96,wakabayashi98,peres2004,fernandez07,fernandez08}, we use a single-orbital Hubbard model to describe $\pi$-magnetism in graphene nanostructures.
The Hubbard model\cite{Arovas2022} can be written as
\begin{equation}
    {\cal H}= \sum_{i,j,\sigma} t_{i,j} c^{\dagger}_{i\sigma}c_{j\sigma} + U\sum_i n_{i\uparrow}n_{i\downarrow},
    \label{eq:Hubbard}
\end{equation}
where the indices $i,j$ run over carbon atoms, $t_{i,j}$ stands for the hopping between sites $i$ and $j$, $U$ is the on-site Hubbard repulsion, $c^{\dagger}_{i\sigma}$ ($c_{i\sigma}$) denotes the operator that creates (annihilates) an electron in site $i$ with spin projection along a quantization axis $\sigma=\uparrow,\downarrow$, and $n_{i\sigma} = c^{\dagger}_{i\sigma}c_{i\sigma}$ is the corresponding number operator. 
While the first term in the Hamiltonian describes hopping between different sites, the second deals with the intra-atomic Coulomb repulsion cost associated to having a given site (or, more formally, the corresponding $p_z$-orbital) doubly occupied.

Unless stated otherwise, we consider systems at half-filling and assume that all $t_{i,j}$ are zero except when $i$ and $j$ are first or third neighbors. 
We denote first and third neighbor hoppings by $t$ and $t_3$, respectively.  
Second-neighbor hopping $t_2$ introduces charge inhomogeneities that are penalized by the Hartree interaction, so that best agreement with DFT is obtained by assuming $t_2=0$.
Throughout this paper we set $t=-2.7$~eV and $t_3$ is taken as a free parameter. 
For triangulene 2D crystals, good agreement with DFT calculations is obtained if we take $t_3\simeq 0.1 t$ (ref.\cite{ortiz22}).

\subsection{CAS}
Due to the exponential increase in complexity as the size of a quantum system grows, exact diagonalization of many-body problems is only possible for rather small systems. 
To treat larger systems, approximate solutions have to be introduced, one of them being the configuration interaction method in the CAS approximation. 
Here, we follow the implementation of the CAS method for the Hubbard model presented in previous works by some of us\cite{ortiz19,Mishra2021,jacob21,jacob22}.
First, the single-particle spectrum of a given triangulene structure is obtained.  
Then, a subset of $N_{\text{MO}}$ molecular orbitals (MOs)---containing the zero modes and closest states in energy---is selected, and a complete set of multi-electronic configurations with $N_e$ electrons occupying these $N_{\text{MO}}$ MOs is considered; the rest of the electrons are assumed to fully occupy the MOs below the active space.  
The Hubbard Hamiltonian is represented in this restricted basis set and diagonalized numerically. 
Hereinafter, we shall refer to this procedure as CAS($N_{\text{MO}}$, $N_{e}$).  
Since there is one electron per $\pi$-orbital for triangulenes at charge neutrality, we always consider $N_e=N_{\text{MO}}$.

\subsection{Mean-field}
In contrast with the CAS method, the mean-field approximation for the Hubbard model makes it possible to include all the single-particle states of molecules and crystals, but interactions are treated approximately.  
The mean-field theory can be formulated variationally, where the many-body wave function is written in terms of a set of independent electrons that occupy the energy levels of a mean-field Hamiltonian. 
Here, we impose that the total $S_z$ is a good quantum number, thus breaking the spin-rotation invariance present in the original Hubbard model; this is the so-called collinear mean-field approximation, extensively used in the modelling of magnetism in graphene nanostructures\cite{fujita96,wakabayashi98,fernandez07,palacios08,fernandez08,yazyev08b,jung09,feldner10,soriano12,ijas13,Ortiz18,zheng20}.
%\cite{yazyev08}
In this case, the Hamiltonian takes the form:
\begin{equation}
\begin{split}
    {\cal H} &= \sum_{i,j,\sigma} t_{i,j} c^{\dagger}_{i\sigma}c_{j\sigma} \\
    &  \quad + U\sum_i \left( \langle n_{i\uparrow}\rangle n_{i\downarrow} + \langle n_{i\downarrow}\rangle n_{i\uparrow} - \langle n_{i\uparrow}\rangle \langle n_{i\downarrow} \rangle \right),
    \label{eq:HubbardMF}
    \end{split}
\end{equation}
where the local densities $\langle n_{i\sigma}\rangle$ are computed with the variational wave function. 
Therefore, the variational wave function and the mean-field Hamiltonian have to be determined in a self-consistent manner. 
In practice, this is done by iteration, starting from an initial guess for the local densities.  
In crystals, the local densities are also periodic so that the eigenvalues and eigenvectors of Eq.~\eqref{eq:HubbardMF} satisfy Bloch's theorem and can be classified in terms of a wave vector $\bm{k}$.

In general, we classify the collinear mean-field solutions in two groups: broken-symmetry solutions for which the expectation values of the {\em local} spin operators are finite, and non-magnetic solutions, that are isomorphic to the non-interacting case, except from a trivial rigid shift of the energies.
Therefore, the mean-field method provides the value of $S_z$ that minimizes the energy, the expectation value of the local moments, and a set of energy levels. 
These three quantities can be compared with DFT. 
In the case of graphene nano-islands\cite{fernandez07} and ribbons\cite{fujita96,son06,fernandez08}, the predictions of mean-field theory were found to be in good agreement with those of DFT for some values of $U$. 
%below the critical value for which 2D graphene is predicted to become antiferromagnetic\cite{sorella92}.
For triangulene 2D crystals we also find a good agreement. 
Therefore, comparison of DFT and mean-field Hubbard models allows us to obtain an educated guess for $U$ in these systems.

In our mean-field calculations for 2D triangulene crystals, we have considered a $5 \times 5$ Monkhorst–Pack grid for the $\bm{k}$-point sums and a tolerance of $10^{-4}$ for convergence in the local densities.
Different initial guesses for the local densities were tested, with the antiferromagnetic guess found to yield the lowest energy solution in all relevant cases.

\subsection{RPA}
\label{sec:methods:RPA}
In order to study spin excitations of 2D triangulene crystals, we use the standard RPA to calculate their transverse spin susceptibility matrix for wave vector $\bm{Q}$ and frequency $\Omega$ (refs.\cite{wakabayashi98,Barbosa2001,peres2004,Culchac2011}),
\begin{equation}
    \chi_{ii'}(\bm{Q},\hbar\Omega) = \frac{1}{N}\sum_{\bm{R}} \mathrm{e}^{\mathrm{i}\bm{Q}\cdot\bm{R}}\int_{-\infty}^\infty \mathrm{d}\mathrm{t} \ \mathrm{e}^{-\mathrm{i}\Omega \mathrm{t}} \chi_{ii'}(\bm{R},\mathrm{t}),
    \label{chiQW}
\end{equation}
which is the space and time Fourier transform of the spin-flip Green function,
\begin{equation}
    \chi_{ii'}(\bm{R}-\bm{R}',\mathrm{t}) = -\mathrm{i}\theta(\mathrm{t}) \left\langle
    [ S^+_{\bm{R},i}(\mathrm{t}), S^-_{\bm{R}',i'}(0) ] \right\rangle,
    \label{chiRt}
\end{equation}
where $i,i'$ are atomic site indices within a unit cell, $\hbar$ is the reduced Planck constant, $N$ is the number of unit cells, $\bm{R},\bm{R}'$ denote unit cell positions, $S^+_{\bm{R},i}(\mathrm{t})$ is the time-dependent version (in the Heisenberg picture) of the spin ladder operator $S^+_{\bm{R},i} \equiv  c^\dagger_{\bm{R},i,\uparrow} c_{\bm{R},i,\downarrow}$ at time $\mathrm{t}$, $S^- \equiv (S^+)^\dagger$, $\theta(\mathrm{t})$ is the unit step function, and $[\cdot,\cdot]$ denotes the commutator. 
The spin-flip Green function depends only on the relative position of unit cells $\bm{R}-\bm{R}'$ due to the translation symmetry of the crystal. 

Within the RPA, we first obtain the mean-field susceptibility $\chi^0$, which corresponds to taking the average $\langle \cdot,\cdot \rangle$ over a self-consistent mean-field state associated with the Hamiltonian given in Eq.~\eqref{eq:HubbardMF}. 
Then the ``interacting'' susceptibility can be calculated using the RPA equation,
\begin{equation}
\begin{split}
\chi_{ii'}(\bm{Q},\hbar\Omega) &= \chi^0_{ii'}(\bm{Q},\hbar\Omega) \\
& \quad - U \sum_{i''} \chi^0_{ii''}(\bm{Q},\hbar\Omega) \chi_{i''i'}(\bm{Q},\hbar\Omega),
\end{split}
\end{equation}
which can be cast in the following matrix form:
\begin{equation}
    [\chi] = (\mathds{1} + U [\chi^0])^{-1} [\chi^0],
\end{equation}
where $[\chi]$ contains the matrix elements $\chi_{ii'}(\bm{Q},\hbar\Omega)$, and analogously for $[\chi^0]$. 
The specific mean-field susceptibilities that are relevant to us are given by Lindhard-like expressions,
\begin{equation}
\begin{split}
\chi^{0}_{ii'}(\bm{Q},\hbar\Omega) = \\ 
\frac{1}{N} \sum_{\bm{k}} \sum_{\lambda,\lambda'} \psi_{\bm{k},\lambda,\up}(i') \psi^*_{\bm{k},\lambda,\up}(i) \psi_{\bm{k}+\bm{Q},\lambda',\dn}(i) \psi^*_{\bm{k}+\bm{Q},\lambda',\dn}(i') \times \\
\times \frac{f(E_{\bm{k},\lambda,\up})-f(E_{\bm{k}+\bm{Q},\lambda',\dn} )}{\hbar\Omega + E_{\bm{k},\lambda,\up} - E_{\bm{k}+\bm{Q},\lambda',\dn} + \mathrm{i}\eta},
\end{split}
\end{equation}
where $\psi_{\bm{k},\lambda,\sigma}(i)$ is the wave function coefficient, at site $i$, of a Bloch eigenstate of band $\lambda$ with wave vector $\bm{k}$ and spin $\sigma$ of the mean-field Hamiltonian. 
The associated eigenvalues are $E_{\bm{k},\lambda,\sigma}$, and $f(E)$ is the Fermi-Dirac distribution function. 
The sum over $\bm{k}$ spans the Brillouin zone of the crystal.
To calculate $\chi^0_{ii'}(\bm{Q},\hbar\Omega)$, we have used 2500 reciprocal space points within the Brillouin zone of the crystal (equivalent to considering $N=2500$ unit cells), which guarantees convergence of the $\bm{k}$-space sum. 
All the results are obtained at zero temperature. 
An empirical broadening of the single-particle states $\eta=5$~meV has been adopted.

The RPA expression for $\chi$ also allows to determine the critical value of $U$, denoted by $U_c$, above which the non-magnetic solutions are no longer stable. 
The magnetic instability is signaled by the condition $\det(\mathds{1} - U_c[\chi^0]) = 0$, with $[\chi^0]$ calculated at $\Omega=0$ for the Hamiltonian in Eq.~\eqref{eq:HubbardMF} with $U=0$. 
The kind of spin arrangement towards which the true self-consistent mean-field solution tends, either ferromagnetic or antiferromagnetic, is indicated by the wave vector $\bm{Q}$ at which the condition is satisfied for the smallest $U_c$, together with the eigenvector of $[\chi^0]$ corresponding to its largest eigenvalue~\cite{MoriyaBook1985}.

\subsection{DFT}
The DFT calculations have been performed with the local-density approximation, as implemented in Quantum Espresso\cite{Giannozzi_2009}. 
We have used norm-conserving pseudopotentials, with a kinetic energy cutoff of 50~Ry and a $\bm{k}$-point sampling of $12\times12\times1$ in a Monkhorst-Pack mesh. 
To avoid spurious interaction between replicas we have set a vacuum distance of 20~\si{\angstrom}. 
We have set the same experimental lattice parameter and atomic positions for the three cases of magnetic order (non-magnetic, ferromagnetic and antiferromagnetic). 
The optimized atomic positions have been calculated using the non-magnetic phase and the final structure is planar.

\subsection{LDOS}
The LDOS at energy $E$ and position $\bm{r} = (x,y,z)$ was calculated using the following equation:
\begin{equation}
    \text{LDOS}(E,\bm{r}) = \sum_{\bm{k}, \lambda, \sigma} 
    | \phi_{\bm{k}, \lambda, \sigma}(\bm{r}) |^2 \delta(E-E_{\bm{k}, \lambda, \sigma}).
    \label{eq:LDOS1}
\end{equation}
The delta function was approximated by a Lorentzian of the form
\begin{equation}
    \delta(E-E_{\bm{k}, \lambda, \sigma}) \simeq \frac{1}{\pi} \frac{\Gamma}{\Gamma^2 + (E-E_{\bm{k}, \lambda, \sigma})^2},
\end{equation}
where $\Gamma$ is the half width at half maximum of the Lorentzian function.
In our calculations, we took $\Gamma = 8$~meV and used a $5 \times 5$ Monkhorst–Pack grid for the $\bm{k}$-point sum.
Moreover, we considered a carbon Slater distribution for the $2p_z$ atomic wave function, 
\begin{equation}
    \phi_{\bm{k}, \lambda, \sigma}(\bm{r}) \propto \sum_{\bm{R}} \mathrm{e}^{\mathrm{i} \bm{k} \cdot \bm{R}} 
    \sum_i \psi_{\bm{k}, \lambda, \sigma}(i) z \mathrm{e}^{- \frac{|\bm{r} - \bm{R}_i|}{r_0}},
    \label{eq:slaterdis}
\end{equation}
with $r_0 = 0.325~\si{\angstrom}$ (refs.\cite{Slater1930,jacob21}).
For clarity, $\bm{R}$ denotes a lattice vector and $\bm{R}_i$ is the specific position of site $i$ in the corresponding unit cell.

%------------------------------------------------------------------------
\section{Hubbard model CAS calculations for centrosymmetric dimers\label{CAS}}
In this section we present the results of CAS calculations for $[n]$triangulene dimers. 
The main goal here is to show that, for $U/|t| \gtrsim 1$, the low-energy spectrum can be mapped to a spin model, providing evidence that the dimers remain open-shell and the triangulenes preserve their magnetic moments. 
%assess the open-shell nature of these molecules, i.e., to determine to what extent intermolecular hybridizations, that could promote closed-shell configurations, are outweighed by Coulomb repulsion. 
%In the range of Hubbard repulsion for which the open-shell nature is established, we determine the parameters for a spin model Hamiltonian describing intermolecular exchange.   
We focus on the cases of $n=4$ and $n=3$ dimers, as the $n=2$ case has been already studied in detail in previous work\cite{jacob22}.  

According to the theorem for the number of zero modes in sublattice-imbalanced bipartite lattices, one should find (at least) $n-1$ zero-energy states for an individual $[n]$triangulene molecule\cite{Sutherland1986,ortiz19}. 
In contrast, $[n]$triangulene dimers have a null sublattice imbalance, so they may have no zero modes. 
However, we find\cite{ortiz22} that there are $2(n-1)$ states close to zero energy, on account of the vanishing weight of the zero modes on the intermolecular binding sites.    
Only third-neighbor hopping leads to a small intermolecular hybridization of the triangulene zero modes\cite{ortiz22}.

\begin{figure}
    \centering
    \includegraphics[width=0.9\linewidth]{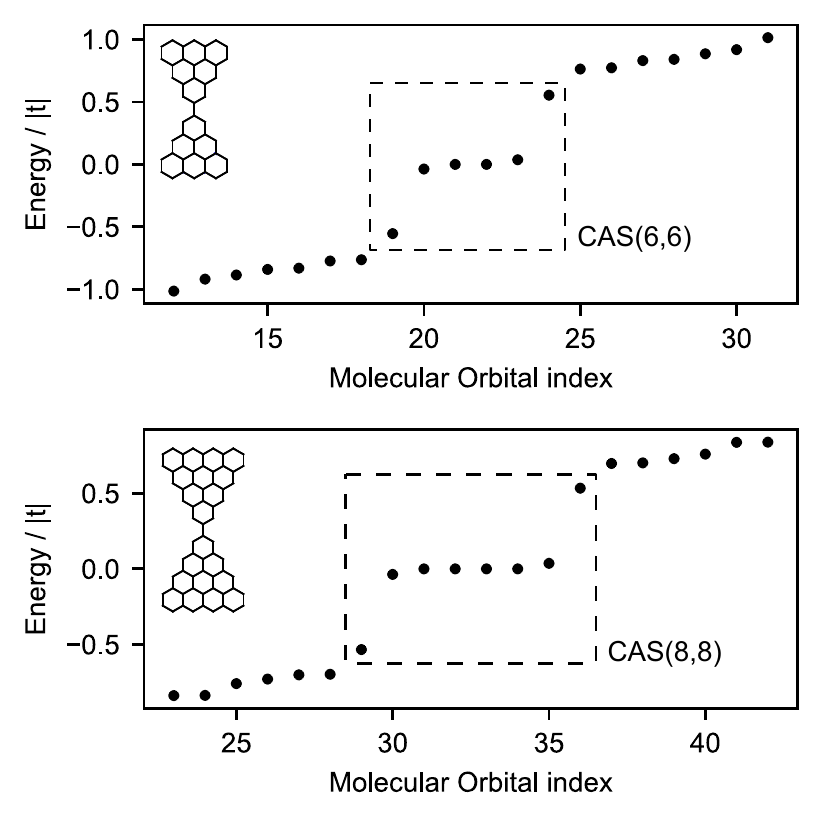}
    \caption{
    Single-particle energy levels of [3]triangulene (top) and [4]triangulene (bottom) dimers, and the respective active space used in the CAS calculations. 
    Results obtained using $t_3 = t/10$.
    }
    \label{fig:Single_Particle_Energy}
\end{figure}

In Fig.~\ref{fig:Single_Particle_Energy} we show the single-particle spectra for [3]- and [4]triangulene dimers, obtained by solving the Hamiltonian of Eq.~\eqref{eq:Hubbard} with $U = 0$ and $t_3 = t/10$. 
As expected, for the [3]triangulene dimer we find four states close to zero energy, originating from the weak intermolecular hybridization of the two zero modes hosted by each monomer individually, promoted by third neighbor hopping. 
For the [4]triangulene dimer, a similar result is found, only this time the three zero modes of the monomers hybridize to give six states close to zero energy. 
We also depict the choice of MOs that will enter in the CAS calculation for each of the molecules. 
These active spaces were chosen to include an additional pair of orbitals besides the zero modes, as this is crucial to account for the Coulomb-driven superexchange mechanism\cite{jacob22}.

\begin{figure}
    \centering
    \includegraphics[width=0.9\linewidth]{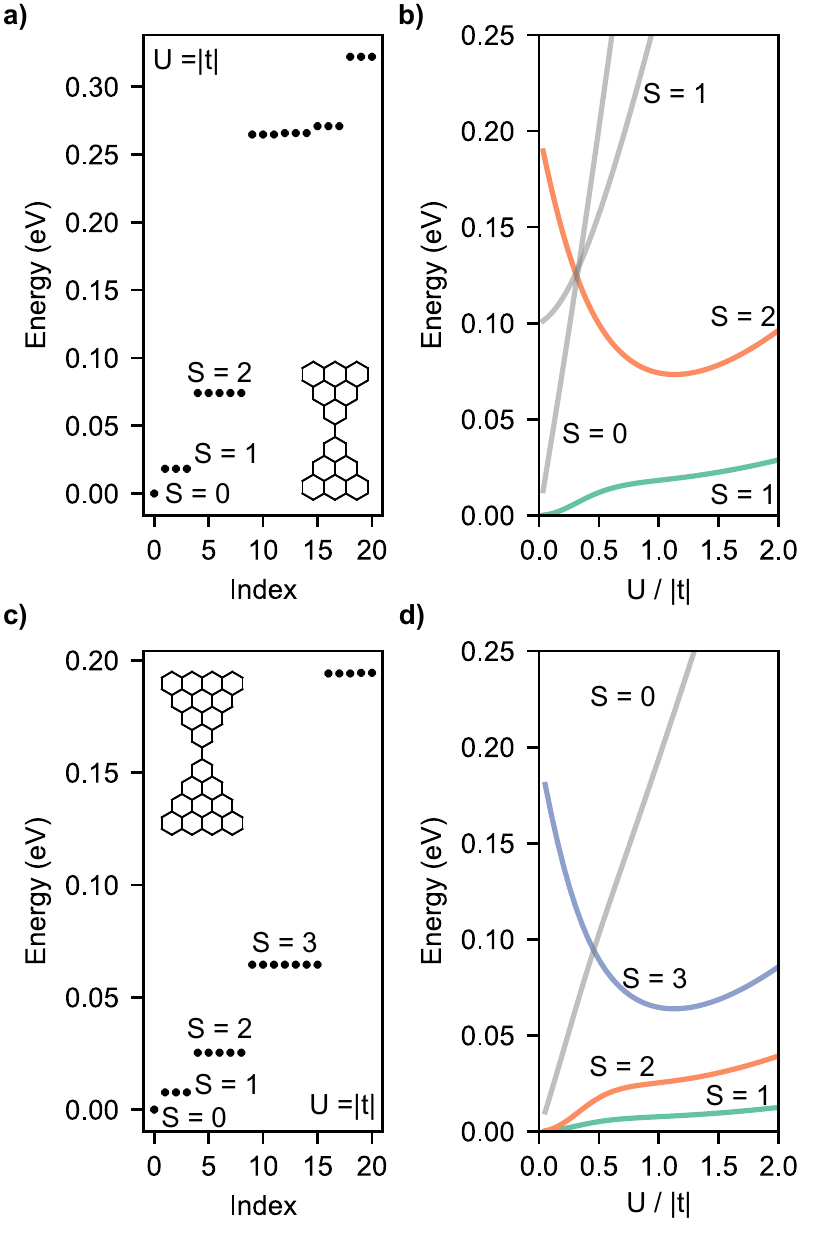}
    \caption{
    Results obtained with CAS for [3]triangulene (a,b) and [4]triangulene (c,d) dimers. 
    Panels (a) and (c) show the energy of the many-body states, obtained with $U=|t|$ and $t_3=t/10$. 
    Panels (b) and (d) show the energy difference between the ground state and the first few excited states, as a function of $U$, for $t_3 = t/10$.
    }
    \label{fig:CAS_Energy}
\end{figure}

We now discuss our CAS calculations for the [3]- and [4]triangulene dimers. 
The results for $U=|t|$ and $t_3 = t/10$ are presented in panels (a) and (c) of Fig.~\ref{fig:CAS_Energy}.
While the [3]- and [4]triangulene monomers are sublattice imbalanced, which according to Lieb's theorem implies a ground state with finite total spin ($S=1$ and $S=3/2$, respectively), for $[n]$triangulene dimers the sublattice imbalance vanishes and the ground state has $S=0$. 
For the $n=3$ dimer, this ground state is followed by a triplet ($S=1$) and a quintet ($S=2$). 
For the $[4]$ triangulene dimer, an additional septet ($S=3$) follows the $S=1$ and $S=2$ manifolds. 
%In both cases,  if sufficiently large ($U \simeq t$ and above), 
%these low lying excitations are well separated from closed-shell $S=0$ excited states, i.e., excitations where two electrons occupy the same molecular orbital. In contrast, for $U<\simeq 0.3t$, the lowest energy closed-shell excitation (with $S=0$) is below the higher energy open shell excitation (with $S=2$ for $n=3$ and $S=3$ for n=4).
%\orangemark{Not true, you can see it for example from Fig 3b. The first state of the high energy manifold is a triplet}  

%

In panels (b) and (d) of Fig.~\ref{fig:CAS_Energy}, we show the CAS results for different values of $U$, thus allowing to study how the energies of the many-body states are affected by the strength of the on-site Coulomb repulsion.
Inspecting this figure, it becomes clear that, for $U \gtrsim |t|$, the low-energy excitation order $S=1$, $S = 2$ (and $S = 3$ for the [4]triangulene dimer) is preserved and, crucially, remains well separated from higher-energy excitations. 
As $U$ is reduced, however, the low-lying excitations become closer to the high-energy ones, and for a critical value of $U$ a crossover is visible. 

In the parameter region where the low-energy manifold is well separated from the higher-energy states, the low-energy spectrum of the triangulene dimers can be modeled by a simple spin Hamiltonian where each triangulene is represented by a spin whose value is that of the ground state of the corresponding monomer. 
To establish a quantitative comparison, we postulate a non-linear Heisenberg dimer Hamiltonian,
%
%\small
\begin{equation}
    H = J \left[ \vec{S}_A\cdot\vec{S}_B + \beta_2 \left(\vec{S}_A\cdot\vec{S}_B\right)^2 + \beta_3 \left(\vec{S}_A\cdot\vec{S}_B\right)^3 \right],
    \label{spin}
\end{equation}
%\normalsize
%
where $\vec{S}_A, \vec{S}_B$ are the vectors of the spin operators for the individual $[n]$triangulenes, taken to be $S_A=S_B \equiv s=1$ and $S_A=S_B \equiv s=3/2$ for $n=3$ and $n=4$, respectively. 

In Appendix~\ref{app:BLBQ}, we derive analytical expressions for the energy levels of this Hamiltonian. 
By matching these expressions with the results found with CAS for the low-energy manifolds of the $[n]$triangulene dimers, we are able to compute $J,\beta_2,\beta_3$ as a function of $U$ and $t_3$. 
As a reference, in Table~\ref{tab:0} we give their values for $U=|t|$ and $t_3 = 0.1t$. 
We see that, for both molecules, the exchange coupling $J$ is in the order of tens of meV, with the $n=3$ dimer presenting the larger antiferromagnetic exchange. 
In both cases, it is found that the biquadratic term (given by $\beta_2 J$) is approximately 10\% of the bilinear one ($J$), emphasizing its importance to accurately capture the energy levels with a spin model. 
As for $\beta_3$, which is only included in the model of the $n=4$ dimer (as explained in Appendix~\ref{app:BLBQ}), it is found to be one order of magnitude smaller than $\beta_2$.
Thus, we see that the bicubic term only introduces minor corrections to the energy spectrum, which further justifies not accounting for it to describe the $n=3$ dimers. 

The fact that we can map the low-energy levels of the fermionic CAS calculation to a spin model, together with the fact that, for $U \gtrsim |t|$, these are well separated from higher-energy excitations provides a strong evidence that the dimers are in the open-shell regime, the triangulenes host local moments, and the singlet ground state arises from the intermolecular antiferromagnetic coupling. 
This shows that, although intermolecular hybridization is present, the magnetic nature of the triangulenes is preserved and the intermolecular interactions are antiferromagnetic. 
A comparison of the intermolecular hybridization and the Coulomb energies is provided in Appendix~\ref{app:openshell}.
%The main take-home message of this section is that, for values of $U \gtrsim t$, the low-energy states of triangulene dimers can be described in terms of spin models. 
As we decrease $U$, the low-energy excitations and the high-energy ones become closer, and the validity of the model is no longer warranted.
The spin model description certainly fails where the crossover between low- and high-energy excitations occurs\cite{Catarina2022}. 

\begin{table}
\begin{tabular}{cccc}
$n$ & $J$ (meV) & $\beta_2$ & $\beta_3$          \\ \hline
3  &   27.9  &   0.12   &   -   \\
4  &   11.3  &   0.09   & 0.007  \\
\end{tabular}
\caption{
Exchange coupling parameters ($J,\beta_2,\beta_3$) obtained by equating the eigenvalues of the non-linear Heisenberg dimer Hamiltonian of Eq.~\eqref{spin} to the CAS results obtained for $[n]$triangulene dimers with $U=|t|$ and $t_3=0.1 t$.
}
\label{tab:0}
\end{table}

%------------------------------------------------------------------------
\section{2D crystals: DFT and mean-field Hubbard model calculations\label{2D}}

In this section we undertake the study of magnetic properties in 2D triangulene crystals. 
For that matter, we compare DFT-based calculations, both spin-restricted and spin-polarized, with mean-field Hubbard model results.  
We consider ferromagnetic (FM) and antiferromagnetic (AF) broken-symmetry solutions, as well as non-magnetic (NM) states.
In all cases considered, we find that the lowest energy configuration corresponds to the AF solution.
%Using both DFT and mean-field Hubbard, we find that the AF state has a lower energy than the FM state. 
%and the non-magnetic (NM) state, whose energy is the highest of the three. 
%\purplemark{This is not true for the [2,2] case at the mean-field level, where I get $E_{FM}-E_{AF}=0.109$~eV and $E_{NM}-E_{AF}=0.097$~eV (for $U=|t|$ and $t_3 = 0.1t$); in ref.\cite{zhou20}, where DFT is done for the [2,2] case, they report $E_{FM}-E_{AF}=0.11$~eV and $E_{NM}-E_{AF}=0.12$~eV. }
%\purplemark{Moreover, I must note that we only have DFT for the [4,4] and for the [3,3] (ref.\cite{ortiz22}) we do not know the energy of the NM solution.}

%
\subsection{DFT for the $[4,4]$triangulene crystal}
We now discuss the electronic properties of the [4,4]triangulene crystal, as described with DFT-based calculations. 
We note that both the spin-restricted and the AF cases of the [2,2]- and  [3,3]triangulene crystals were addressed in previous works\cite{zhou20,ortiz22}.  
In both systems, it was found that the NM solution is an excited state and describes a zero-gap semiconductor with two Dirac cones and a narrow bandwidth.  
The spin-polarized AF solution opens up a large gap 
%(0.716 for AF phase, 0.380 for FM phase), 
and is the ground state.
%\purplemark{About the numbers: ref.\cite{zhou20} ([2,2] case) does not give the number, but looking at the figure it should be around 600~meV; same for the [3,3] case (ref.\cite{ortiz22}), with the gap being around 800~meV.}

Previous work\cite{sethi2021,ortiz22,delgado23} has shown that the spin-unpolarized [4,4]triangulene crystal is a narrow-gap semiconductor with flat valence and conduction bands.   
Here, we go beyond the NM framework and study two magnetic phases, AF and FM. 
We find that the AF phase has smaller energy than both the FM ($E_{FM}-E_{AF}=0.171$~eV) and the NM ($E_{NM}-E_{AF}=0.457$~eV).
It is thus apparent that DFT calculations confirm the open-shell nature of the [4]triangulenes when covalently bonded to form a 2D honeycomb crystal.
If we model the energy difference between the AF and FM phases with a classical Heisenberg model on a honeycomb lattice, we get
%the magnetic energy difference per unit cell between the FM and AF solutions is given by 
$6JS^2=0.171$~eV. 
Using $S=3/2$, we pull out $J=12.7$~meV.
For the [3,3]triangulene crystal, a similar analysis\cite{ortiz22} found $E^{[3,3]}_{FM}-E^{[3,3]}_{AF}=0.159$~eV and $J^{[3,3]}=26.5$~meV. 
%In Table~\ref{tab:1} (and \ref{tab:J}) we compare these numbers with the predictions of the mean-field Hubbard model.

\begin{figure}
    \centering
    \includegraphics[width=\linewidth]{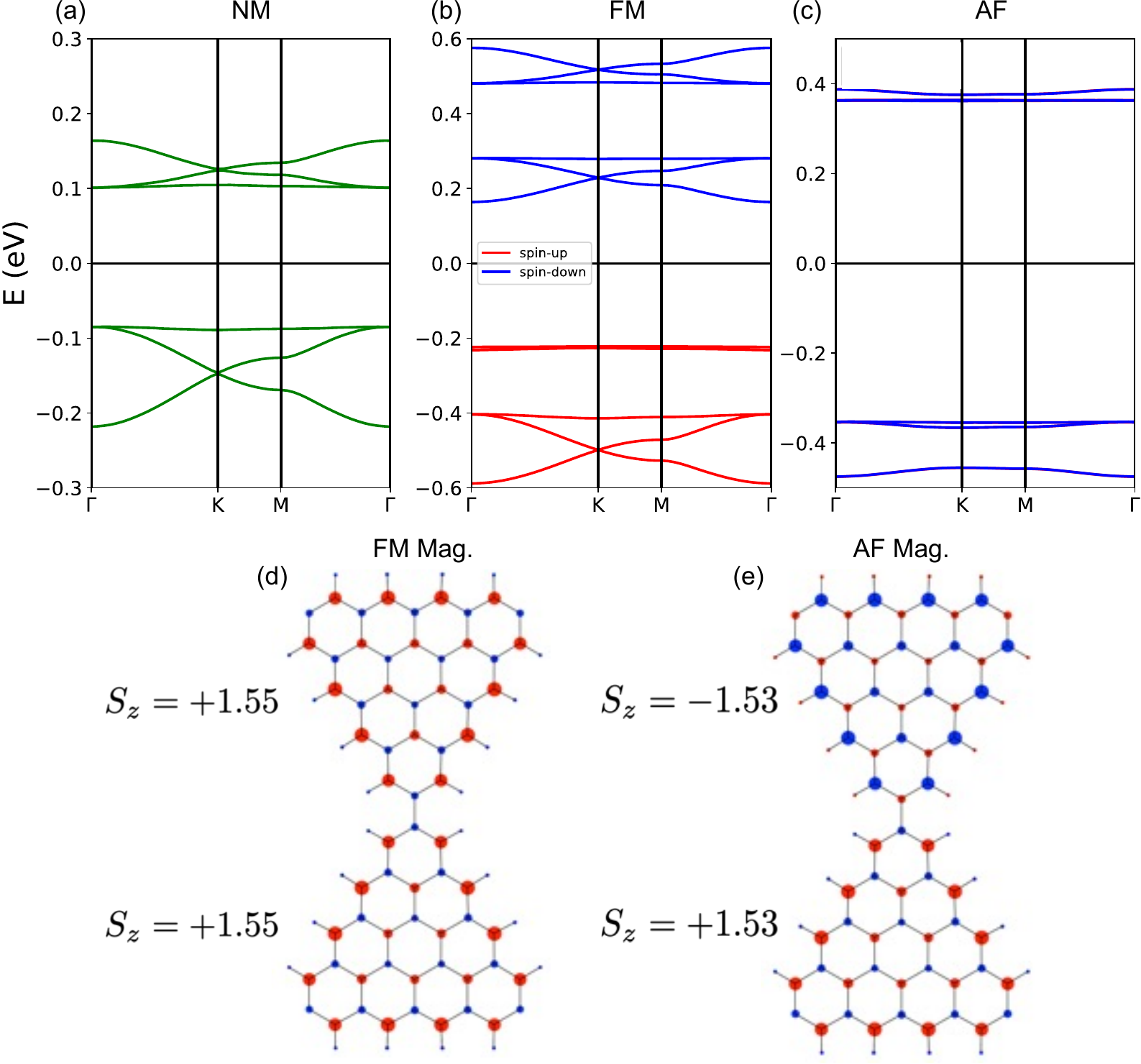}
    \caption{
    Electronic structure, obtained with DFT, of [4,4]triangulene 2D crystals for the following cases: (a) non-magnetic, (b) ferromagnetic and (c) antiferromagnetic.
    The corresponding magnetizations of the FM and AF cases are shown in panels (d) and (e), respectively; the value of $S_z$ per triangulene is also indicated.
    Red/blue colors correspond to spin-up/down. 
    In (c), spin-up and spin-down bands are degenerate.
    }
    \label{fig:DFT}
\end{figure}

In Fig.~\ref{fig:DFT}, we show the energy bands for the three configurations (NM, FM, AF) of the [4,4]triangulene crystal, together with the distribution of the magnetic moments in the FM and AF solutions.  
The three solutions are gapped, but the size of the gap increases in the magnetic phases, specially in the AF case.  
The FM bands have a similar line shape than the NM bands, except for the top of the conduction band.
The AF bands are much narrower than the NM bands.  
%As we discuss below, 
This relates to the quenching of intermolecular hybridization due to the opposite-sign spin splitting of the zero modes of adjacent molecules.
%, that competes with intermolecular hopping. 

We note that, whereas the magnetic moments lie predominantly in the majority sublattice of each triangulene, there is a smaller magnetization with opposite sign in the minority sublattice, coming presumably from electrons in non-zero modes. 
Moreover, we find that the magnetization per triangulene shares a similar pattern for both FM and AF solutions, and the values obtained are compatible with the predictions for individual triangulenes.

\subsection{Mean-field Hubbard model results}

We now present our results for the [2,2]-, [2,3]-, [3,3]-, and [4,4]triangulene 2D crystals, obtained using the collinear mean-field approximation to the Hubbard model at half-filling.  
For the centrosymmetric $[n,n]$trianguelene crystals, we find that, for $U$ above an $n$-dependent critical value $U_c (n)$ below which the ground state solutions are NM (see Subsection~\ref{sec:uc}), the lowest energy solutions are AF, in agreement with DFT calculations.
As for the non-centrosymmetric [2,3] case, the ground state obtained is always ferrimagnetic.

\begin{figure*}
    \centering
    \includegraphics[width=\linewidth]{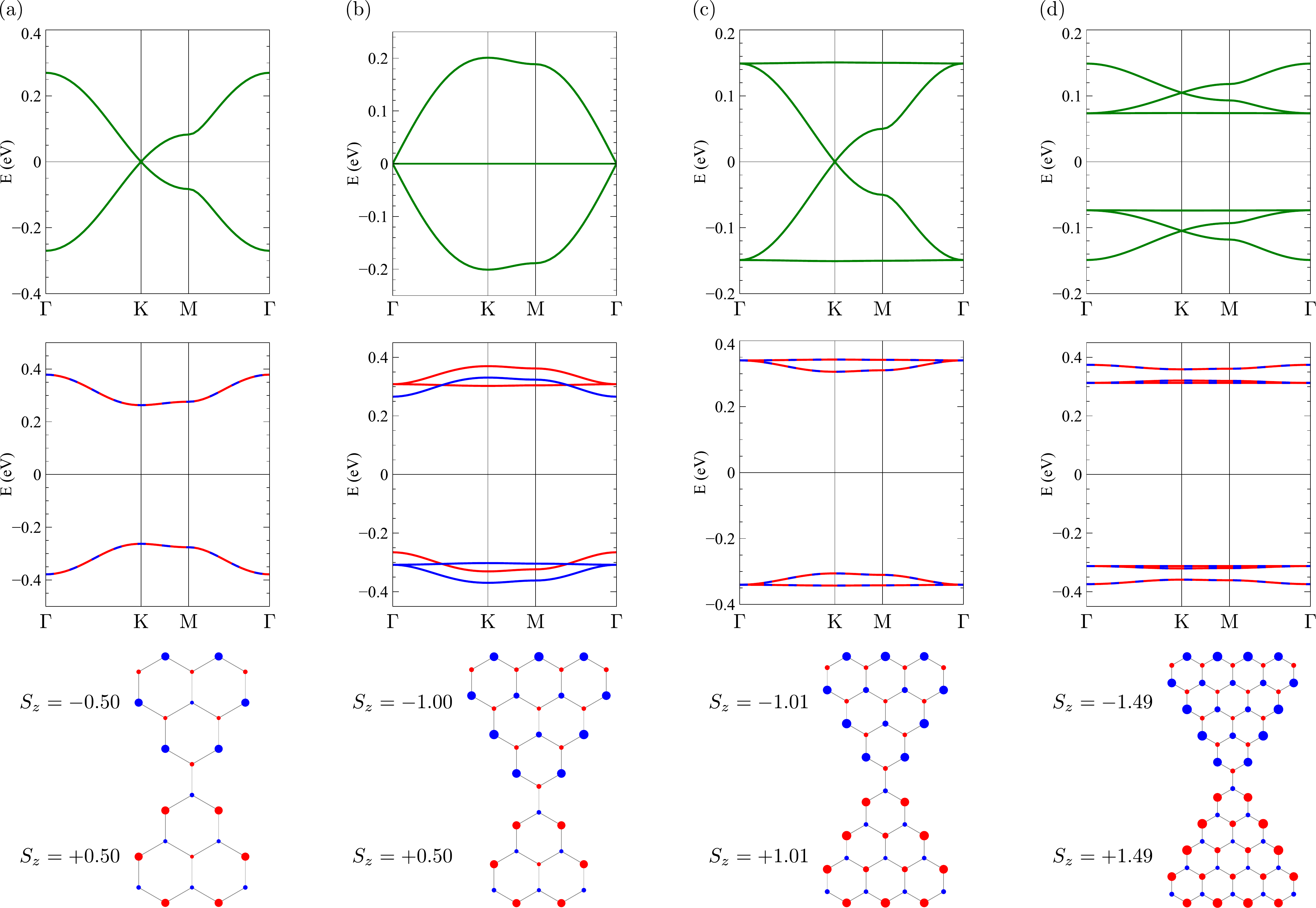}
    \caption{
    Electronic properties of (a) [2,2]-, (b) [2,3]-, (c) [3,3]- and (d) [4,4]triangulene 2D crystals.
    Top panels show the NM energy bands, obtained using a tight-binding model with $t_3 = 0.1t$; horizontal black lines denote the Fermi energy.
    Middle panels show the energy bands of the ground state solution of a collinear mean-field Hubbard model with $U = |t|$; red/blue colors denote spin-up/down.
    The corresponding magnetizations are shown in the bottom panels, where the size of the circles represents the magnitude of the local moments.
    The value of $S_z$ per triangulene is also indicated.
    }
    \label{fig:5}
\end{figure*}

In Fig.~\ref{fig:5}, we show the energy bands for both NM and ground state (magnetic) configurations, obtained with $U=0$ and $U=|t|$, respectively.
Two features are immediately apparent. 
First, the dispersion of the AF bands is narrower compared to the NM case.
This is a consequence of suppressed intermolecular hybridization, on account of the opposite-sign spin splitting in the two triangulenes of the unit cell.  
Second, the separation between conduction and valence bands increases in the magnetic phases. 
Thus, the [2,2]-, [2,3]-, and [3,3]triangulene crystals, gapless for $U=0$, become gapped when magnetic order appears. 
In the case of the [4,4]triangulene crystal, gapped for $U=0$, the interactions increase the gap by more than a factor of 3. %(see also Table~\ref{tab:1}). 
The gap of the magnetically ordered phases reflects the fact that every triangulene is full-shell in a spin-channel, so that the addition of a new electron is only possible in the minority spin channel, that became spin-split. 

In Fig.~\ref{fig:5}, we also show the local magnetic moments corresponding to the ground state mean-field solutions.
The magnetization pattern is such that carbon sites in different sublattices are magnetized with opposite sign.  
For $U \simeq |t|$, the magnetic moments per triangulene are close to the values expected from Lieb's theorem for individual triangulenes, and in qualitative agreement with those of DFT. %(see comparison in Table~\ref{tab:1}). 
We note that mean-field theory is not constrained by Lieb's theorem, that applies to exact solutions. 

For the non-centrosymmetric [2,3]triangulene crystal, the magnetic order appears for arbitrarily small values of $U$. 
This is expected on account of the flat band at the Fermi energy. 
%Magnetic order is ferrimagnetic. 
For small values of $U$, magnetic moments are only present in the larger unit, that hosts the flat-band states. 
As $U$ is ramped up, the magnitude of the magnetic moments in both units increases towards values close to those of the isolated triangulenes, and a ferrimagnetic ground state is obtained. 

\subsection{Comparison between mean-field and DFT \label{sec:mfvsdft}}
In this section, we briefly compare the results of the mean-field Hubbard models with those of DFT, for the $[2,2]$, $[3,3]$ and $[4,4]$ crystals.
The DFT results for the $[2,2]$ crystals are taken from ref.\cite{zhou20}.
As for the [3,3] crystals, DFT calculations were reported in ref.\cite{ortiz22} by two of us.
Since the comparison of the NM phases was already established in previous work\cite{ortiz22}, we focus on the magnetic phases.
 
Qualitatively, both levels of theory are in agreement.
They both predict AF solutions as the ground state, with magnetic moments close to those predicted for isolated triangulenes. 
Moreover,  both in mean-field and DFT the band-gap of the magnetic solutions is much larger than the NM cases, and the band dispersion is narrower.

Given the uncertainty over the best value of $U$, we make no attempt to find the value of $U$ for which this agreement is better, and we take $U=|t|$ as a reasonable guess.
%, which is a good representative of the values found in the literature. 
It is apparent that the mean-field bands obtained with $U=|t|$ are in good agreement with the DFT calculations.
The same is also verified for the magnetization patterns (see figures (\ref{fig:DFT})d,e and lower panels (\ref{fig:5}).
A quantitative comparison between the mean-field theory for $U=|t|$ and DFT is provided in Table~\ref{tab:1}.
We find a fairly good agreement that justifies the use of Hubbard models for this type of system. Specifically, for the $[4,4]$ case, 
%Remarkably, we find that, for the [4,4]triangulene crystal, 
we obtain a good agreement in: (i) the energy difference between the different magnetic phases (with the NM configuration featuring the highest energy of the three); (ii) the band gaps of the NM and AF solutions, with both levels of theory predicting an increase of the band gap by a similar factor in the AF phase; (iii) the $S_z$ per triangulene (discussed above); (iv) the absolute value of the magnetization, defined by $|M_\text{tot}| = g \mu_\text{B} \sum_i |S_z(i)|$, where $g=2$ is the electron $g$-factor and $\mu_\text{B}$ stands for the Bohr magneton.

\begin{table}
    \begin{tabular}{cccc}
System & Quantity & DFT & Mean-field \\ \hline
$[2,2]$ & $E_{FM}-E_{AF}$ (eV) & 0.11$^a$ & 0.109 \\
$[2,2]$ & $E_{NM}-E_{AF}$ (eV) & 0.12$^a$ & 0.097 \\
$[3,3]$ & $E_{FM}-E_{AF}$ (eV) & 0.159$^b$ & 0.137 \\
$[4,4]$ & $E_{FM}-E_{AF}$ (eV) & 0.171 & 0.133 \\
$[4,4]$ & $E_{NM}-E_{AF}$ (eV) & 0.457 & 0.508 \\
NM $[4,4]$ & Gap (eV) & 0.185 & 0.148 \\
AF $[4,4]$ & Gap (eV) & 0.716 & 0.625 \\
AF $[4,4]$ & $|M_{\rm tot}|$ ($\mu_\text{B}$) & 8.89 & 9.01 \\
AF $[4,4]$ & $S_z$ per triangulene & 1.53 &  1.49     
\end{tabular}
\caption{
Agreement between DFT and mean-field Hubbard model calculations for different magnetic phases of $[n,n]$triangulene 2D crystals.
Mean-field results were obtained with $t_3 = 0.1t$ and $U = |t|$.
$^a$: Ref.\cite{zhou20}; $^b$: Ref.\cite{ortiz22}.
}
\label{tab:1}
\end{table}

\subsection{Critical value of $U$
\label{sec:uc}}
\begin{figure}
    \centering
    \includegraphics[width=0.9\columnwidth]{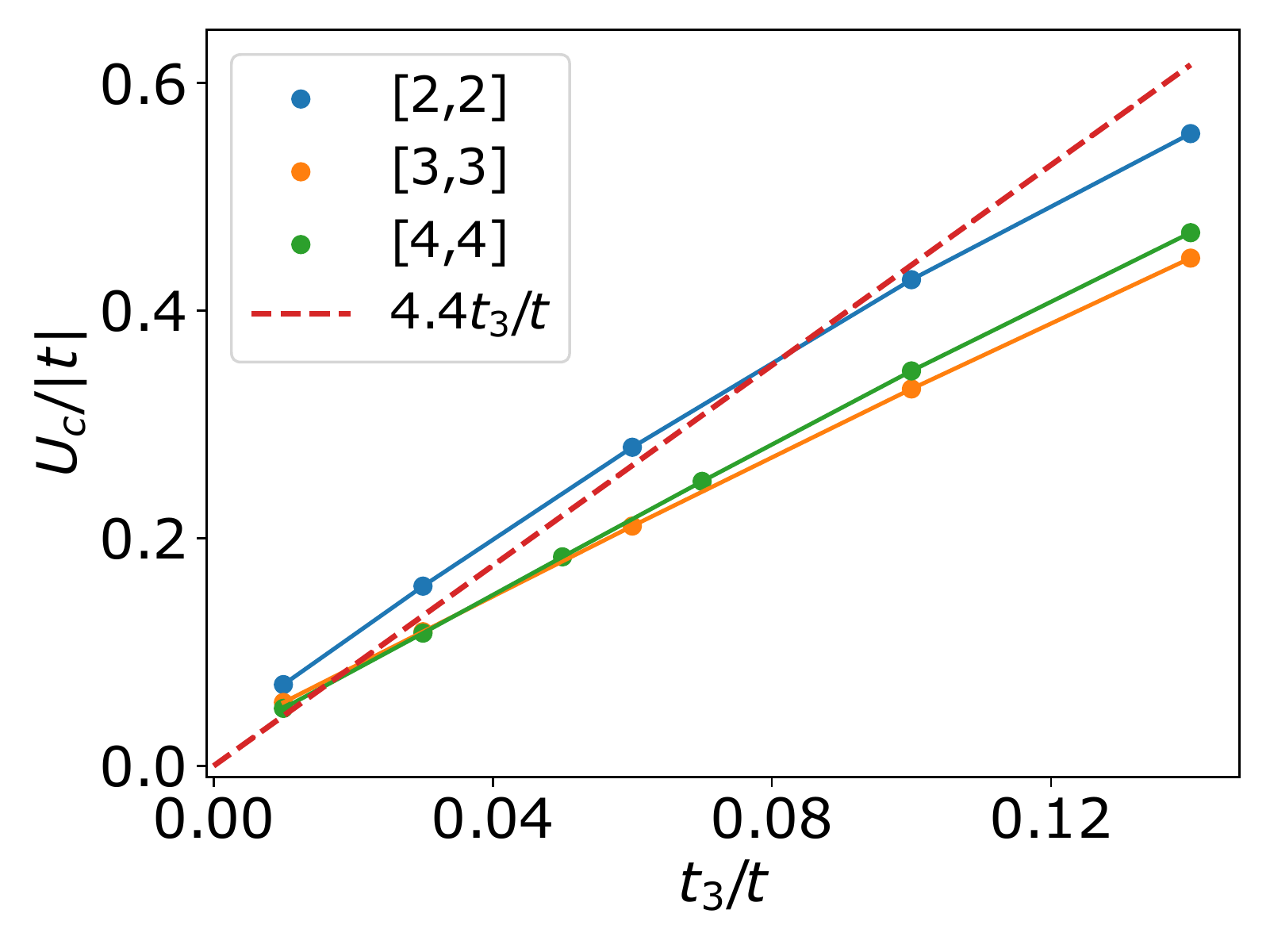}
    \caption{
    %\redmark{Changes: 1. Change units of vertical axis to $U/t$.  2.  Change labels to $[2,2]$, $[3,3]$ $[4,4]$ } 
    Critical values $U_c$ of the Hubbard parameter for the onset of the magnetic instability,
    as predicted by the RPA, as a function of the third neighbor hopping $t_3$, for the [2,2]-, 
    [3,3]- and [4,4]triangulene crystals.}
    \label{fig:Uc_x_t3_x_n}
\end{figure}

We now discuss the minimal value of $U$ that makes the NM solution unstable within the mean-field Hubbard approximation.  
This can be obtained in two ways. 
First, by comparing the NM and the magnetic solutions of a mean-field calculation as a function of $U$ and finding the critical value $U_c$ above which the disordered phase becomes an excited state.  
Second, a faster approach, discussed in Section~\ref{sec:methods:RPA} and adopted here, where we look for the value $U=U_c$ for which the non-interacting RPA susceptibility diverges.  
The results are shown in Fig.~\ref{fig:Uc_x_t3_x_n} for the $[n,n]$ crystals with $n=2,3,4$.  
We note that, for a honeycomb Hubbard model at half-filling, the mean-field critical value for the NM to AF transition is $U_c=2.2 |\tau|$ (ref.~\cite{sorella92}), where $\tau$ is the first neighbor hopping of the honeycomb lattice. 

% Comparison with Sorella's Uc=2.2 t
For the $[2,2]$triangulene crystal, whose low-energy single-particle Hamiltonian maps exactly to that of a honeycomb model\cite{ortiz22}, the effective first neighbor hopping is given by $\tilde{t}=|t_3|/3$ and the effective Hubbard interaction is $\tilde{U}=U/6$ (ref.\cite{ortiz22}). 
Therefore, 
%we obtain $\frac{\tilde{U}}{\tau}= \frac{U/6}{|t_3|/3}=2.2$, from which 
by renormalizing the $U_c=2.2 |\tau|$ equation,
we can estimate a critical value $U_c=4.4 |t_3|$ for the [2,2] crystal, in good agreement with the $U_c$ numerically obtained (see Fig.~\ref{fig:Uc_x_t3_x_n}). %\redmark{Antonio, check how does this compare with your numbers.  From eye-inspection, your results for $n=2$ are $U_c\simeq \frac{0.75}{0.06} \frac{t_3}{t} [eV]\simeq 4.6 t_3$, so, this looks good }.

 Given that the low-energy spectrum of the $[3,3]$triangulene crystal also features graphene-like bands in the neighborhood of the Fermi energy,  we can also compare the numerically obtained $U_c$ with that of the honeycomb crystal.
 For the $[3,3]$ crystal, the  effective first neighbor hopping is given by $\tilde{t}=2 |t_3|/11$ and the effective Hubbard is $\tilde{U}\simeq U/11$ (ref.\cite{ortiz22}). %\orangemark{are we sure about this? I think the number is $35U/365$}. 
 Therefore, we 
 %get $\frac{\tilde{U}}{\tau}\simeq \frac{U/11}{2 |t_3|/11}=2.2$, from which we 
 also estimate a critical value $U_c\simeq 4.4 |t_3|$.  
 The fact that our numerical estimates for $U_c(t_3)$ are slightly different (see Fig.~\ref{fig:Uc_x_t3_x_n}) reflects the fact that $U_c$ is also influenced by the flat bands away from the Fermi energy.

The critical values of $U$ for $t_3=0.1 t$ are in the range of $U_c \lesssim 0.45 |t| < 1.2$~eV. Estimates of atomic $U$ for carbon are higher than this, in the range of 3.5~eV (ref.\cite{yazyev10}). 
Mean-field theories are known to underestimate $U_c$. 
For instance, Quantum Monte Carlo methods\cite{sorella92} predict $U_c=4.5 |\tau|$ for the Hubbard model on the honeycomb lattice.  
Even if $U_c$ is twice as large as the values  predicted by mean-field,  magnetic order should appear in the triangulene crystals. 

Interestingly, the value of $U_c$ is very similar for the $[3,3]$ and $[4,4]$ crystals.  
This result further supports the picture that, once moderately large interactions are included, the  fact that the NM bands of the $[4,4]$ crystal have a band-gap does not seem to have a dramatic effect on its electronic properties, and the $[4,4]$triangulene crystal is (antiferro) magnetic.

\section{Collective spin excitations in 2D triangulene crystals\label{RPA}}

%\begin{itemize}
  %  \item Broken spin-rotational symmetry implies the existence of low energy Goldstone modes
 %   \item The poles of the RPA spin susceptibility permit us to compute them (cite previous work by Antonio, also Nuno's paper, and some classics)
%    \item We can compute poles frequency as a function of wave-vector $\omega(q)$.  We find two (\redmark{right?} degenerate bands, shown in the figure. 
 %   \item They are identical/similar to the band-dispersion of a Heisenberg model with $J=$ . Compare this $J$ with $J_{CAS}$ and $J_{DFT}$.
  %  \item In principle, spin-excitations of both chains (cite STM papers) and 2D crystals (cite my paper with Pablo Jarillo) can be seen with IETS. 
 %   \item Spin waves are interesting on their own right, as they can be used to transfer information, etc. 
%\end{itemize}

%\redmark{Joaquin's comment for Antonio: not sure about whether the $T=0$ ground state preserves symmetry. It may break it. At least, it can have long range order}.
%\bluemark{Although the ``true'' ground state of a 2D isotropic antiferromagnet is 
%hitherto unknown, it is informative to look at the properties of spin excitations 
%above the broken-symmetry state characterized by N\'eel order.}
\subsection{RPA for the Hubbard model}
The choice of a ``ground state'' with broken spin rotation symmetry implies the 
existence of gapless Goldstone modes, the spin waves. Here we obtain the spin wave 
spectra of 2D triangulene lattices by computing the transverse spin susceptibility 
$\chi(\boldsymbol{Q},\hbar\Omega)$ of the Hubbard Hamiltonian (Eq.~\eqref{eq:Hubbard}) in the RPA, as discussed in Section~\ref{sec:methods:RPA}. The spin 
wave frequencies are associated with the poles of $\chi(\boldsymbol{Q},\hbar\Omega)$. 
For a given wave vector, two poles occur at energies $\pm\hbar\Omega(\boldsymbol{Q})$, 
due to the opposite directions of the spins in the two magnetic sublattices.\cite{Fisher1971}
From those we can build a spin wave dispersion relation, shown in Figs.~\ref{fig:RPA4}a,b for the $[3,3]$ and  $[4,4]$ crystals, and in Fig.~\ref{fig:RPA4}c for the $[2,3]$ crystal.
We note that, for the centrosymmetric
cases ($[3,3],[4,4]$), the two modes are degenerate, in contrast with the $[2,3]$ for which we find an acoustic and an optical branch of spin waves. In these figures,  
 the symbols represent the 
locations of the poles of $\chi(\boldsymbol{Q},\hbar\Omega)$ for a few wave vectors along
two high-symmetry directions in the honeycomb Brillouin zone. It is apparent that the bandwidth of the magnon spectrum
is larger for the $[3,3]$ crystal, in agreement with the larger values of intermolecular exchange obtained with the CAS calculations 
for the dimers.

\subsection{Comparison with spin models}
We now compare the RPA results with those of a Heisenberg spin model with first-neighbor exchange $J$, calculated  in the linear spin-wave approximation\cite{holstein40,auerbach98}.  The calculation (not shown) is standard\cite{auerbach98}.  The spin operators are expressed in terms of Holstein-Primakoff (HP) bosons\cite{holstein40}, taking the quantization axis parallel to the classical ground state (AF for the $[3,3],[4,4]$, ferrimagnet for the $[2,3]$), where the classical magnetization of each $[n]$triangulene is $2S=n-1$. The resulting bosonic Hamiltonian is truncated so that only  terms bilinear in the HP bosons are kept.  This bilinear Hamiltonian can be solved exactly, by means of a paraunitary canonical transformation.  For a lattice with two spins per unit cell, such as the honeycomb, two spin-wave branches are obtained, given by:
%\orangemark{
\begin{eqnarray}
\frac{\epsilon_{\pm}(\boldsymbol{k})}{3J}= \frac{S_A+S_B}{2}\sqrt{1- \xi_{\boldsymbol{k}}}\pm\frac{S_B-S_A}{2},
\label{eq:SW}
\end{eqnarray}
where $\xi_{\boldsymbol{k}} = 4 S_A S_B |\phi_{\boldsymbol{k}}|^2 /(S_A + S_B)^2 $, 
$S_A$ and $S_B$ denote the spin of the triangulenes in sublattice $A$ and $B$, $3\phi_{\boldsymbol{k}}=1+\mathrm{e}^{\mathrm{i}\boldsymbol{a}_1\cdot\boldsymbol{k}}+\mathrm{e}^{\mathrm{i}\boldsymbol{a}_2\cdot\boldsymbol{k}}$, $\boldsymbol{a}_{1,2}$ are the lattice vectors of the honeycomb lattice and $J$ is the intermolecular exchange.  It is apparent that in the AF case we have $S_A=S_B$ and the two branches become degenerate. It is also apparent that, for $\boldsymbol{k}=(0,0)$, the lower energy branch vanishes, complying with the Goldstone theorem.

\begin{figure}
    \centering
    \includegraphics[width=\columnwidth]{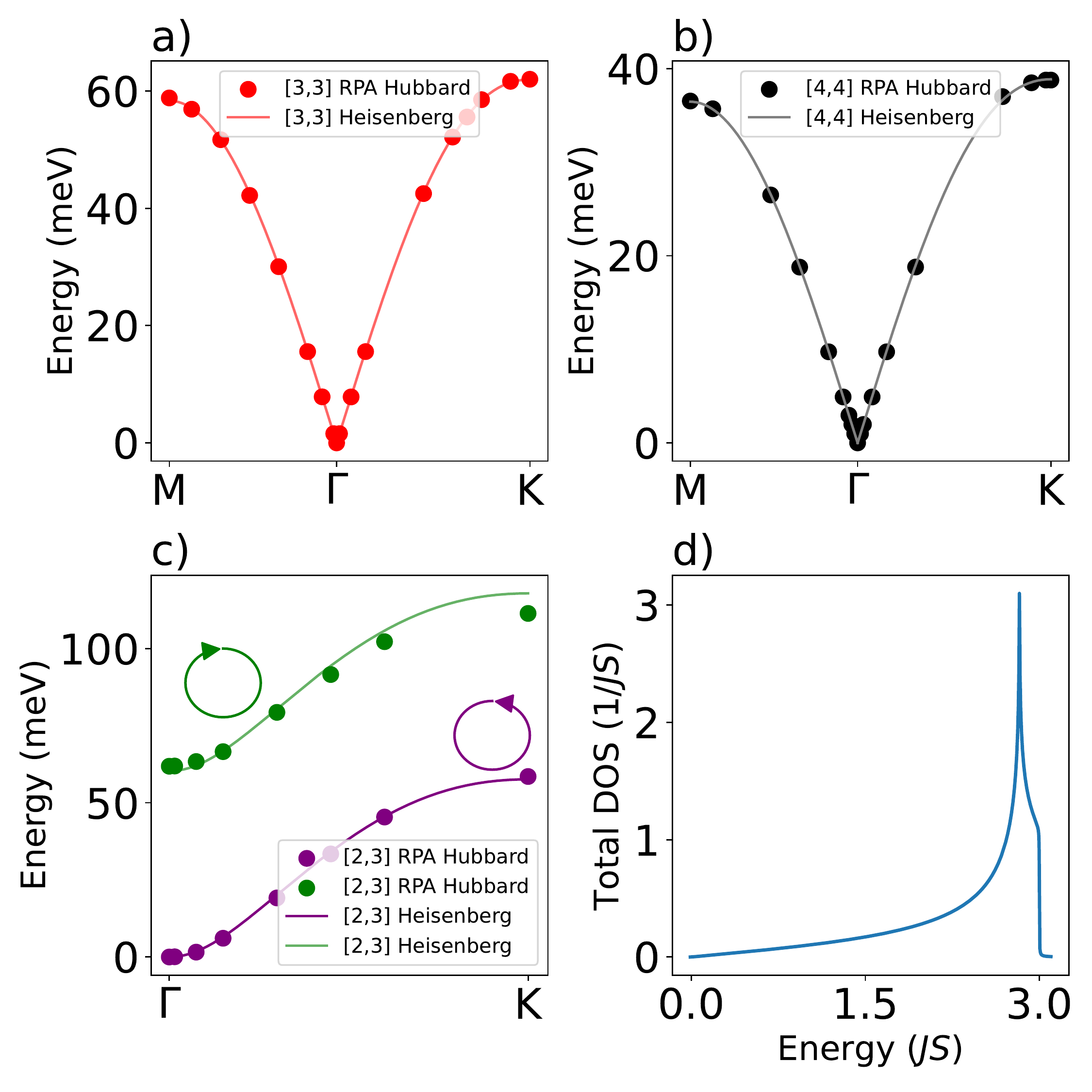}
    \caption{Spin wave dispersion relations within the RPA for $[n,m]$triangulene crystals. a) [3,3] crystal, b) [4,4] crystal, c) [2,3] crystal, for which the two spin wave branches (corresponding to different polarities) are non-degenerate due to the ferrimagnetic nature of the ground state. In contrast, for the $[3,3]$ and $[4,4]$ (antiferromagnetic) crystals, the two polarities are degenerate, thus a single dispersion is shown for each. In a), b) and c) panels, the dots have been extracted from magnon spectral densities (the imaginary part of the transverse spin susceptibility) and the solid curves are fits to nearest-neighbor Heisenberg models. d) Integrated magnon density of states for the nearest-neighbor antiferromagnetic Heisenberg model on the honeycomb lattice, obtained from Eq.~(\ref{eq:SW}) with $S_A=S_B=S$.} %\redmark{Still ugly, I know. I'll make it prettier.}    
    %Effective exchange as a function of $U$, obtained by fitting the spin wave dispersion relation to a nearest-neighbor Heisenberg model in the linear spin wave approximation (dots), and from CAS calculations on single dimers (solid curves).}
    \label{fig:RPA4}
\end{figure}

In Figs.~\ref{fig:RPA4}a--c, we   compare the magnon dispersion calculated from the fermionic RPA theory with the  spin wave dispersion of Eq.~(\ref{eq:SW}) .
Taking $S$ from the mean-field calculation, we determine the value of intermolecular exchange $J$ that provides the best fitting to the RPA calculation within the fermionic model. We find that the RPA curves lies
exactly 
on top of the spin-wave curves, providing additional support to the notion that the low-energy exciations of 2D triangulene crystals  can be  described with spin model Hamiltonians, very much like the one-dimensional triangulene spin chain.
%From this fitting, and the value of $S$ provided by the self-consistent
%mean-field determination of the N\'eel configuration, we can extract a value for the

\begin{figure}
    \centering
    \includegraphics[width=0.9\columnwidth]{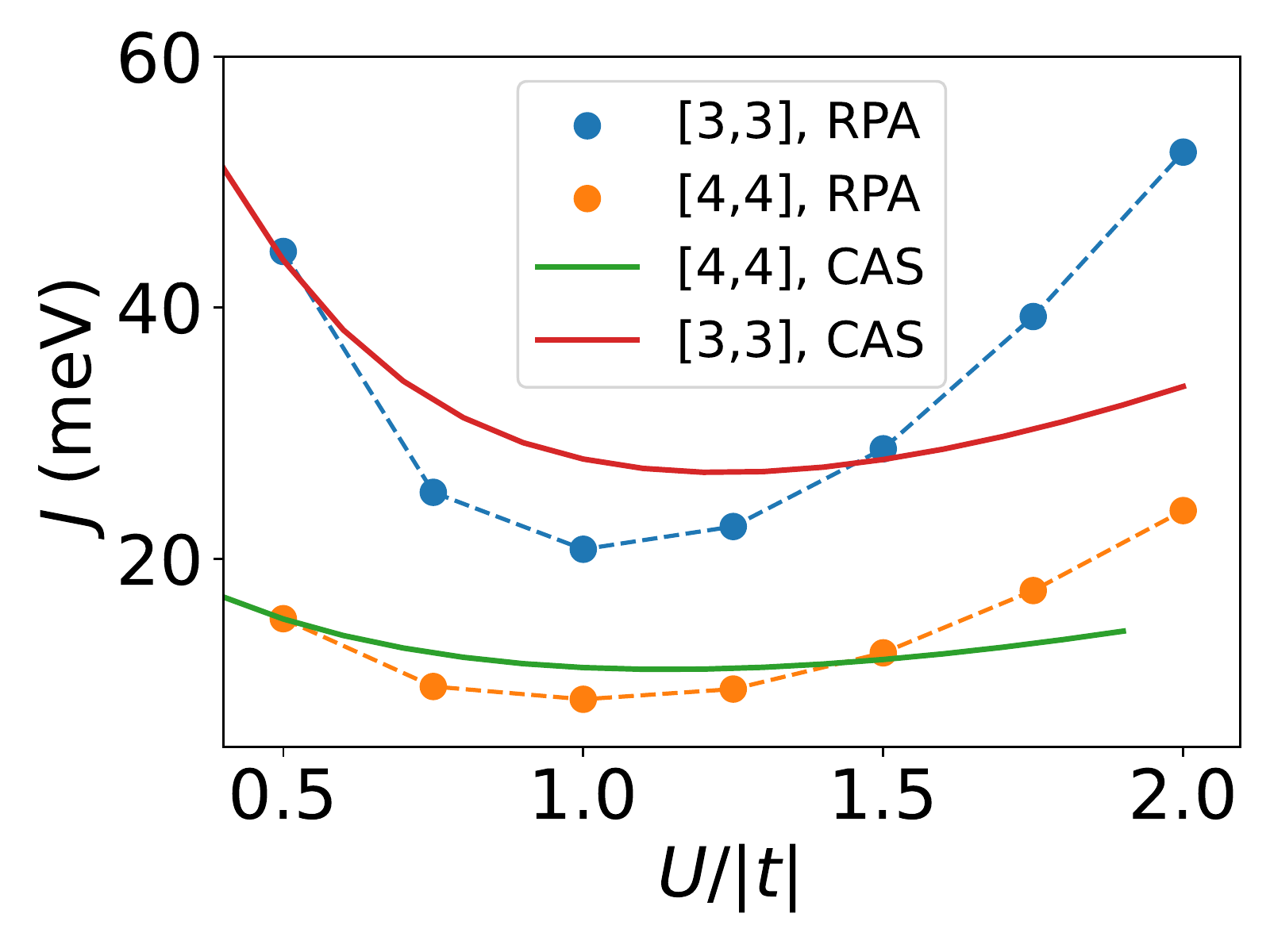}
    \caption{Effective exchange as a function of $U$, obtained by fitting the spin wave dispersion relation to a nearest-neighbor Heisenberg model in the linear spin wave approximation (dots), and from CAS calculations on single dimers (solid curves).}
    \label{fig:JxURPA}
\end{figure}

We can determine the dependence of $J$ on $U$ and $t_3$ by repeating this 
procedure for different values of those parameters.
In Fig.~\ref{fig:JxURPA} we plot $J$, so obtained, as a 
function of $U/|t|$ with $t_3=t/10$ for the $[3,3]$ and $[4,4]$ cases. The general 
behavior is qualitatively very similar to the results from CAS calculations for dimers,
as seen in Figs.~\ref{fig:BLBQBC}a,c. In fact, even the actual values of $J$ given by
RPA and CAS are reasonably similar for $0.5 \lesssim U/|t|   \lesssim 1.5$.  This qualitative good agreement backs-up the robustness of the main underlying picture of this work: despite the intermolecular hybridization between triangulenes  in the  2D crystals considered here, they retain their magnetic moment.

%\begin{table}
 %   \begin{tabular}{ccccc}
 %DFT        &  Mean Field &  CAS& RPA & system  \\ \hline
 %26.5 (meV) & 22.2 meV 
 %&  27.9 meV (\redmark{check}) 
 %&  \redmark{fill} 
 %& $[3,3]$  \\
%12 meV &   19 meV  
% &  11.3 meV (\redmark{check})
% &  \redmark{fill}
% & $[4,4]$
%\end{tabular}
%\caption{
%Estimates for intermolecular exchange from DFT and from Hubbard model, taking $U=|t|=2.7 eV$, and $t_3=0.1 t$, in three different approximations 
%}
%\label{tab:J}
%\end{table}

\begin{table}
    \begin{tabular}{c|cccc|c}
 \hspace{0.5cm} &DFT        &  Mean-field &  CAS& RPA  \\ \hline
$J^{[3,3]}$ 
(meV) & 26.5 $^a$  & 22.8
 &  27.9   
 &  20.8 
 %& $[3,3]$  
 \\
$J^{[4,4]}$ (meV) & 12.7 &   9.9
 &  11.3  
 &  8.8
 %& $[4,4]$
\end{tabular}
\caption{
Estimates for intermolecular exchange from DFT and from Hubbard model, in three different approximations,  taking $U=|t|$, and $t_3=0.1 t$, for the [3,3] and [4,4] triangulenes.
 $^a$: Ref.\cite{ortiz22}.
%\orangemark{Very small detail, but we should be consistent with the number of decimal places}
}
\label{tab:J}
\end{table}

\section{Predictions for STM spectroscopy \label{LDOS}}

We now discuss experimental consequences of the magnetic order discussed in the previous sections. Given that, so far, triangulenes structures are studied with STM\cite{mishra2021b,hieulle2021,delgado23}, we focus properties that can be probed with this technique.  STM $dI/dV$ can reveal two different properties of the surface\cite{Ortiz2020}: LDOS  and inelastic excitations.  In the case of nanographenes, LDOS features are revealed as prominent peaks at large voltages, in the range of hundreds of meV,  corresponding to resonant tunneling accross  specific energy levels of the molecules.  

\subsection{Probing  LDOS }
Here we discuss the LDOS at the energy of the valence and conduction bands, that can be measured by means of STM spectroscopy.  The LDOS is sensitive to the interatomic coherence: by virtue of Eq.~(\ref{eq:LDOS1}), the LDOS is proportional to the square of the MO wave function, that in turn is a linear combination of atomic orbitals. Therefore,  LDOS is sensitive to the relative phases of the weights of the MO at different atoms.   Specifically, for the non-interacting bands of a bipartite lattice,  electron-hole symmetric states, such as  valence and conduction bands have opposite relative phases between adjacent atoms. More formally, let us denote  the wave function of a conduction band MO as
\begin{equation}
\psi_{\bm{k}, \lambda}(i)=
\left(\psi_{\bm{k}, \lambda}(A), \psi_{\bm{k}, \lambda, \sigma}(B)\right) ,   
\label{eq:13}
\end{equation}
where 
$\psi_{\bm{k}, \lambda, \sigma}(A/B)$
encodes the MO weight on all the atoms in the unit cell that belong to the $A$ and $B$ sublattices. Then, for a bipartite lattice, the wave function of the electron-hole conjugate state $\overline{\lambda}$ in the valence band  is given by\cite{soriano12}
\begin{equation}
\psi_{\bm{k}, \overline{\lambda}}(i)=
\left(\psi_{\bm{k}, \lambda}(A), -\psi_{\bm{k}, \lambda, \sigma}(B)\right)
\label{eq:14}
\end{equation}

Thus, electron-hole conjugate MO wave functions have the same probability {\em amplitudes} but opposite {\em phases} at one sublattice. As a result, LDOS will have a enhancement/depletion at the regions connecting atoms with different sublattices. 
Specifically, at the bonding region between any pair of atoms, we can truncate Eq.~(\ref{eq:slaterdis}) keeping only  contribution of the
two closest atoms, $a$ and $b$, that, by definition, 
belong to different sublattices. This leads to
%\begin{equation}
 %   \phi_{\bm{k}, \lambda}(\bm{r}) \simeq 
  % \sum_{i=A,B} \left(\psi_{\bm{k}, \lambda}(i) |\bm{r} - \bm{R}_i| \mathrm{e}^{- \frac{|\bm{r} - \bm{R}_i|}{r_0}}\right)
   %\label{eq:approxslater}
%\end{equation}
\begin{equation}
    \phi_{\bm{k}, \lambda}(\bm{r}) \simeq 
   \left[\psi_{\bm{k}, \lambda}(a)
   g(\bm{r}-\bm{R}_{a})+
  \psi_{\bm{k}, \lambda}(b)
   g(\bm{r}-\bm{R}_{b})\right]
   \label{eq:approxslater}
\end{equation}
where $g(\bm{r}-\bm{R}_{A/B}) \equiv z \mathrm{e}^{- \frac{|\bm{r} - \bm{R}_{A/B}|}{r_0}}$. %$f(\bm{r}-\bm{R}_{A/B})=|\bm{r} - \bm{R}_{A/B}| \mathrm{e}^{- \frac{|\bm{r} - \bm{R}_{A/B}|}{r_0}}$.

We can now compute the {\em difference} of the LDOS computed in the bonding region between two atoms, for which eq. (\ref{eq:approxslater}) holds, evaluated at energies $+E_\lambda$ and $-E_\lambda$.  
The contributions to the {\em difference} of LDOS at these two energies from states with the same wave vector $\boldsymbol{k}$ in a pair of electron-hole conjugate bands $\lambda,\overline{\lambda}$ will be given by
%\begin{equation}
%    \delta \rho(\lambda,\boldsymbol{k},\bm{r}) = 4
%\psi_{\bm{k}, \lambda}(A) 
%\psi_{\bm{k}, \lambda}(B)
%    f(\bm{r}-\bm{R}_A)  f(\bm{r}-\bm{R}_B)
%     \label{eq:contrast}
%\end{equation}
%\orangemark{
\begin{equation}
    \delta \rho_{\boldsymbol{k},\lambda}(\bm{r}) = 4 \text{Re}\left[
\psi_{\bm{k}, \lambda}^*(a) 
\psi_{\bm{k}, \lambda}(b) \right]
    g(\bm{r}-\bm{R}_a)  g(\bm{r}-\bm{R}_b)
     \label{eq:contrast}
\end{equation}
%}
From Eq.~(\ref{eq:contrast}) it is apparent that the LDOS contrast between electron-hole bands is controlled by the weights (and crucially the respective phases) of the wave functions of the MOs
of adjacent atoms  (which belong to different sublattices). %For instance, for sublattice polarized modes, the contrast of eq. (\ref{eq:contrast} vanishes.

\begin{figure}
    \centering
    \includegraphics[width=\linewidth]{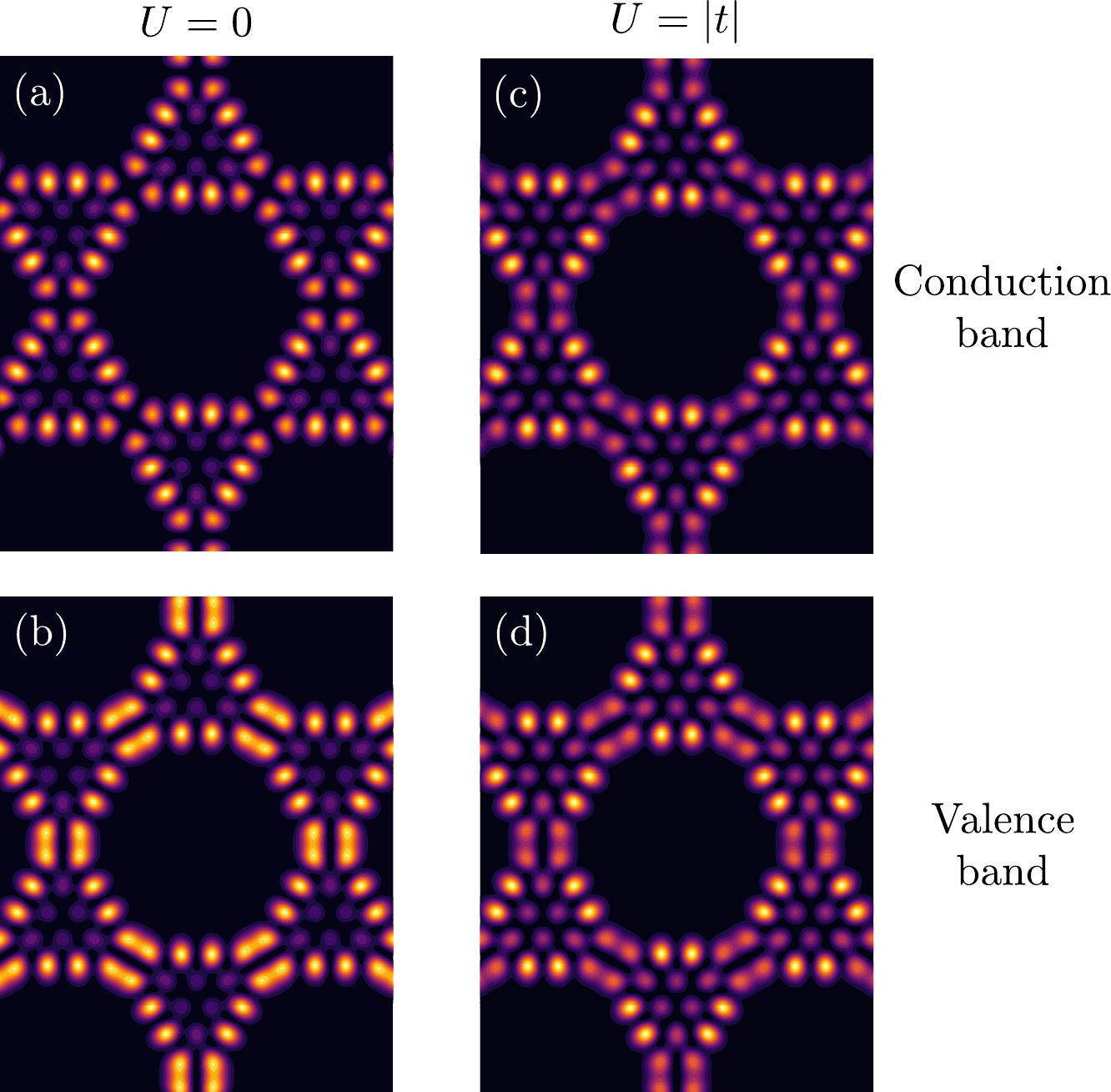}
    \caption{ 
    LDOS for [4,4]triangulene 2D crystals.
    All calculations were performed at height $z=2.8$~\si{\angstrom}.
    Panels (a,b) show the results obtained with the non-interacting tight-binding model, with $t_3 = 0.1t$, for energies in (a) conduction band ($E=0.1$~eV) and (b) valence band ($E=-0.1$~eV).
    Panels (c,d) show the results obtained with the collinear mean-field Hubbard model, taking $U = |t|$, for energies in (c) conduction band ($E=0.315$~eV) and (d) valence band ($E=-0.315$~eV).
    }
    \label{fig:LDOS}
\end{figure}

In the case of conduction and valence bands of triangulenes, the weight of the wave functions inside each triangulene is all on a single sublattice. Therefore, the LDOS contrast is only seen at the inter-triangulene  binding sites. These are shown for the $[4,4]$ triangulene crystal in Fig. \ref{fig:LDOS}. For the non-interacting case we  find   a depletion of the LDOS at the intermolecular  binding sites  at the conduction band energy and a corresponding enhancement of the valence band (see Fig. \ref{fig:LDOS}a,b). 

We now discuss how interactions, described at the mean-field level,  change this picture. The broken-symmetry N\'eel states result in the presence of a staggered exchange potential. As a result, for a given spin direction, the on-site energy of two adjacent triangulenes is no longer the same. Consequently, the wave functions of valence and conduction bands no longer have the same weight on both sublattices \cite{soriano12}, i.e.  Eqs.~(\ref{eq:13}) and (\ref{eq:14}) relating the wave functions of valence and conduction bands no longer hold.  In the interacting cases the MOs become sublattice biased, and in the very strong coupling limit, completely sublattice polarized. This  ultimately reduces the amplitude of the bonding-anti-bonding interference effect, as shown in Fig.~\ref{fig:LDOS}c,d . This is in agreement with the experimental observations of Delgado et al.\cite{delgado23}.

We note here that the reduction of the LDOS contrast between valence and conduction bands in the interacting cases relates to the  reduced bandwidth of the interacting bands. The spin-dependent staggered potential creates an energy barrier for intermolecular hybridization.

In the work of Delgado et al., \cite{delgado23}, the observed reduced contrast of LDOS at the binding site is  attributed to an excitonic insulator state that arises on account of the small gap obtained from the spin-unpolarized  DFT calculations. As both our  DFT  and mean-field results show, the spin polarized solution has lower energy, a larger gap (that makes the excitonic insulator state less likely) and, more important, already accounts for the reduced LDOS contrast in terms of the sublattice symmetry breaking of the AF solution. 

%Heuristically,  the magnetization-driven  sublattice localization can be recasted in terms of a mixing of the non-interacting valence and conduction band, akin to an excitonic insulator.  When a only a single pair of bands is involved in the process, as in the case of zizgzag ribbons,  there is a formal mapping between mean-field  theory for magnetic zigzag edges and an excitonic insulator\cite{fernandez08}. However, we note that this scenario is different from the one proposed by Delgado and coworkers\cite{delgado23}, where the band-mixings is attributed to an excitonic insultor that develops over a non-magnetic ground state. 

\subsection{Probing magnons}
 In contrast to the large-bias LDOS measurements discussed above,  STM spectroscopy can reveal  inelastic excitations  as bias-symmetric steps, at bias voltages below 100~meV, whose height is dramatically smaller than the resonant peaks. The underlying mechanism for these inelastic steps is  inelastic cotunneling of electrons\cite{delgado11,Ortiz2020}. In spin systems, inelastic electron tunneling spectroscopy (IETS) can probe spin transitions between the ground state and excited states that satisfy the rule for the change of total spin $\Delta S=0,\pm 1 $. Therefore,  IETS is optimal to probe magnons\cite{spinelli14,klein18}.  We expect that $\frac{d^2I}{dV^2}$ will have a line shape that reflects the density of states of magnon excitations. In Fig.~\ref{fig:RPA4}d, we show the magnon LDOS associated to  the dispersion energy from Eq.~(\ref{eq:SW}) for $S_A=S_B=S$, relevant for  $[n,n]$triangulene crystals.  It is apparent that the magnon LDOS features an outstanding Van Hove singularity,  at the energy $\sqrt{8}JS$, corresponding to the $M$ points in the Brillouin zone.  Therefore, we anticipate the presence of a prominent feature at $eV=\pm \sqrt{8}JS$  energy in the $\frac{d^2I}{dV^2}$ spectra.   For the $[4,4]$triangulene crystal, taking $J\simeq 9$~meV (see Table~\ref{tab:J}), and $S=3/2$,  the steps are expected at $\pm 38$~meV.

\section{Summary and conclusions\label{discuss}}

The main goal of this paper is to describe the consequences of electron-electron interactions in  triangulene 2D crystals. Specifically, we address the question of  whether triangulenes retain their magnetic moments when forming 2D crystals that entail intermolecular hybridization.   This is particularly relevant in the case of $[4,4]$ triangulene crystals, for which the single-particle model\cite{ortiz22} predicts an insulating state that, naively, may quench the emergence of magnetism. 

We employ both spin-unrestricted DFT calculations and Hubbard model with three different approximations: 
 \begin{enumerate}
 \item  Multiconfigurational calculations of triangulene dimers. %showing ground states with antiferromagnetically coupled triangulene spins

 \item Mean-field approximation of the 2D crystals, whose results are in qualitative agreement with DFT calculations. % predicting AF coupled triangulenes

 \item RPA calculations of the spin excitations. %
 \end{enumerate}

Importantly, these different methods  permit us to perform cross-validations. For instance, both the  spin-polarized DFT and mean-field Hubbard model yield very similar results for all the key quantities (see table \ref{tab:1}).  The Hubbard-model  RPA calculations, building on top of mean-field solutions,  predict an excitation spectra that can be fitted very well to a Heisenberg model, with just a single fitting parameter, the effective exchange (see figure (\ref{fig:RPA4})a,b,c).  In turn, this exchange  is in qualitative agreement with the one obtained from CAS calculations, for a range of values of $U$ (see figure (\ref{fig:JxURPA})).

Our main conclusions are:
\begin{enumerate}

\item Triangulenes retain their magnetic moment when forming in two-dimensional crystals, according to   both DFT and Hubbard model calculations.

\item  Triangulene crystals are insulating, on account of the electron-electron interactions. This is supported both by our mean-field Hubbard model and our spin-polarized DFT calculations.  In the case of the $[4,4]$ crystal, the size of the gap, calculated with DFT, comes out 3.9 times  larger than the spin-unpolarized  gap, which  calls for a revision of the  predictions\cite{sethi2021,delgado23} of an excitonic insulator state based on the smaller gap of the non-magnetic ground state. 

\item Two-dimensional triangulene crystals are  magnetically ordered, either antiferrromagnetically, in the  centrosymmetric case, or ferrimagnetically, for non-centrosymmetric crystals.  This statement is based on both DFT and mean-field Hubbard calculations.  

\item The value $U=|t|$ gives a very good agreement between mean-field Hubbard model and DFT  for  
several quantities, such as the intermolecular exchange, the magnetic moment and the band-gap. We have not tried to fine-tune $U/|t|$ to improve that agreement, but we can be sure  that $U\simeq |t|$ is a good ball-park reference for this important ratio, in agreement with previous work\cite{mishra2021b}.

\item The low energy spin excitations obtained from the RPA fermionic calculations are very well described with Heisenberg Hamiltonians (see Fig. \ref{fig:RPA4}). The  exchange interactions so obtained are in qualitative agreement with those obtained from CAS, mean-field and DFT (see table \ref{tab:J}). 

\item Intermolecular exchange features non-linear interactions, beyond Heisenberg. This is found by comparing
 CAS calculations for the Hubbard model with spin models.  For $U\simeq |t|$, the values of the non-linear terms are in qualitative agreement with previous work for $S=1$\cite{mishra2021b}.  For $S=3/2$ triangulenes, the value of the non-linear interactions are very small, so that it is very unlikely that the system realizes the AKLT model for the honeycomb lattice \cite{Affleck1987}, that would be for relevant measurement-based quantum computing \cite{Wei2011}. 

\item Magnetically ordered  states reduce the intermolecular hybridization in triangulene 2D crystals. This has two consequences: 
\begin{itemize}
    \item First,  the bandwith of magnetically ordered triangulene crystals are narrower than those of the non-interacting case. 
    \item Second, 
a reduction in the difference of the LDOS measured at the valence and conduction band energies at the intermolecular binding sites.   This reduction has been observed experimentally in recent work \cite{delgado23} in $[4,4]$ triangulene crystals.  
\end{itemize}  

 \end{enumerate}

We now briefly discuss the robustness of the predicted antiferromagnetic states. 
 By construction,  both the  mean-field and DFT calculations predict broken-symmetry solutions or non-magnetic solutions.  Both quantum and thermal fluctuations can destroy long range order in two dimensions. 
 
 At $T=0$,  broken symmetry states are robust in this class of system. Using Quantum Monte Carlo, it was shown that the Hubbard model in the honeycomb lattice, at half-filling,  features antiferromagnetic  long-range order\cite{sorella92}. Since the effective model for the $[2,2]$ crystal is a Hubbard Hamiltonian that maps into a $S=1/2$ Heisenberg model,  given that quantum fluctuations scale with $1/S$, and given that larger triangulenes have larger $S$, we expect that at $T=0$ the ground state of the centrosymmetric triangulene crystals also  feature  N\'eel long-range order.   
 
In contrast, thermal fluctuations are expected to destroy long-range order, on account of  Mermin-Wagner theorem\cite{mermin66}. However, the spin correlation length may be larger than system size for small crystals reported experimentally\cite{delgado23}. Therefore, the broken-symmetry solutions remain a good approximation for these systems, as in the case of one-dimensional edge magnetism in graphene ribbons\cite{yazyev08}. 

Our results, together with previous experimental work\cite{Mishra2021,delgado23}, should pave the way for the design of other nanographene molecular crystals\cite{ortiz22, ortiz2023a,ortiz2023b}, both 1D and 2D that realize interesting spin Hamiltonians with non-trivial electronic properties.     

\vspace{1cm}

\section*{Acknowledgements}

We acknowledge discussions with David Jacob and Ricardo Ortiz-Cano. 
G.C. acknowledges financial support from Funda\c{c}\~{a}o para a Ci\^{e}ncia e a Tecnologia (FCT) for the PhD scholarship grant with reference No. SFRH/BD/138806/2018.
J.F.R., J.C.G.H. and A.C  acknowledge financial support from 
%1
 FCT (Grant No. PTDC/FIS-MAC/2045/2021),
 %2
 SNF Sinergia (Grant Pimag),
 %3
 the European Union (Grant FUNLAYERS
- 101079184).
J. F. R. acknowledges funding from
FEDER /Junta de Andaluc\'ia, %--- Consejer\'ia de Transformaci\'on Econ\'omica, Industria, Conocimiento y Universidades,
(Grant No. P18-FR-4834), 
% 4
Generalitat Valenciana (Prometeo2021/017
and MFA/2022/045)
%5
and
MICIN-Spain (Grants No. PID2019-109539GB-C41 and PRTR-C1y.I1) 
%"This study forms part of the Advanced Materials programme and was supported by MCIN with funding from European Union NextGenerationEU (PRTR-C17.I1) and by Generalitat Valenciana (MFA/2022/045)."
A.M.-S. acknowledges financial support by Ram\'on y Cajal programme (grant RYC2018-024024-I; MINECO, Spain), Agencia Estatal de Investigación (AEI), through the project PID2020-112507GB-I00 (Novel quantum states in heterostructures of 2D materials), and Generalitat Valenciana, program SEJIGENT (reference 2021/034), project Magnons in magnetic 2D materials for a novel electronics (2D MAGNONICS). This study forms part of the Advanced Materials programme and was supported by MCIN with funding from European Union NextGenerationEU (PRTR-C17.I1) and by Generalitat Valenciana, project SPINO2D, reference MFA/2022/009.

\appendix

\section{Derivation of spin Hamiltonian parameters
\label{app:BLBQ}}

\begin{figure}
    \centering
    \includegraphics[width=\linewidth]{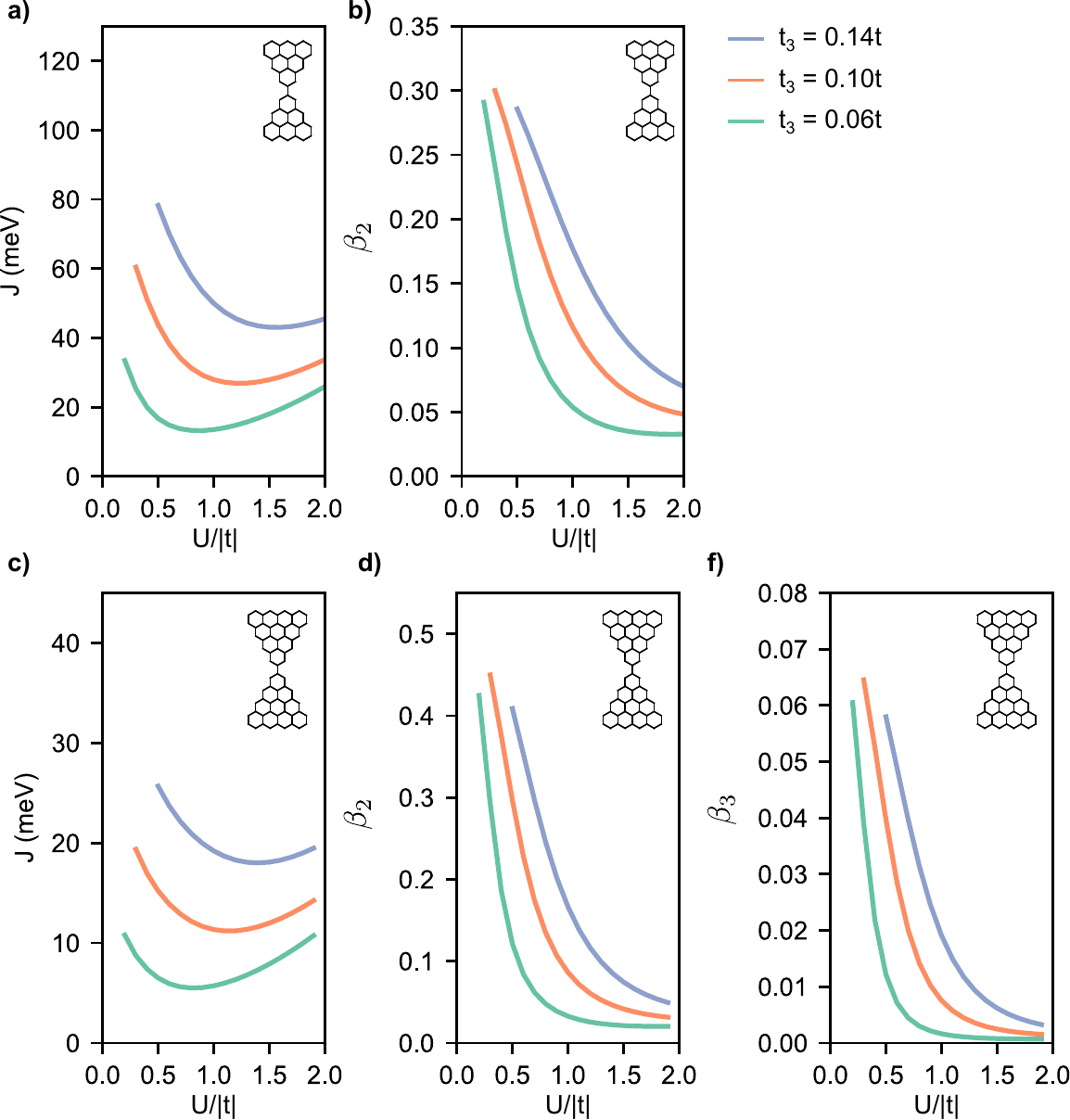}
    \caption{ 
    Parameters of the spin model, obtained by matching the spin dimer energy levels to the CAS calculations for triangulene dimers, as a function of $U$, for $t_3 = 0.06t$, $t_3 = 0.1t$ and $t_3 = 0.14t$. 
    %The dashed lines mark the value of $U$ for which the crossover between low- and high-energy excitations occurs for each $t_3$. 
    Panels (a) and (b) show the results for the [3]triangulene dimer, and panels (c)-(e) show the results for the [4]triangulene dimer.
    }
    \label{fig:BLBQBC}
\end{figure}

In this Appendix, we derive the analytical expressions for the energy levels of the spin model dimer Hamiltonian of Eq.~\eqref{spin}. 
Then, by matching these expressions with the numerical results obtained with CAS for $[n]$triangulene dimers, we study how the parameters of the spin model, i.e. $J$, $\beta_2$ and $\beta_3$, depend on $U$ and $t_3$.

In terms of the total spin $S$ and the spin of the triangulenes $s$, the eigenvalues of the spin model of Eq.~\eqref{spin} are given by
\begin{equation}
      E_s(S)=J\left[f(S,s)+ \beta_2 f(S,s)^2  + \beta_3 f(S,s)^3 \right],
      \label{spin2}
\end{equation}
with
\begin{equation}
    f(S,s)=\frac{1}{2}\left[S(S+1)- 2s(s+1)\right],
\end{equation}
where $S$ can take the values $S=0, 1, ...,  2s$.  
Thus, for the $n=3$ ($n=4$) case, $S$ can take values up to $S=2$ ($S=3$). 
For the [4]triangulene dimer, the spectrum of the spin model has four multiplets, with $S=0,1,2,3$.  
For the [3]triangulene dimer, we assume $\beta_3=0$ since the $S=1$ dimer model can take the values $S=0,1,2$, and we can only fit two energy parameters out of three multiplets. 
As we found in previous work\cite{Mishra2021}, the model with $\beta_3=0$ can account for experimental observations of a large number of structures.  

%Taking $\beta_3=0$, we obtain the following relation between the excitation energies and the $J,\beta_2$ parameters:
%\orangemark{
The excitation energies for the $n=3$ case are related to $J$ and $\beta_2$ as follows:
\begin{eqnarray}
E_1(1)-E_1(0)=J(1-3\beta_2), \\
E_1(2)-E_1(0)=3J(1-\beta_2). 
\end{eqnarray}
These equations can be easily inverted to obtain $J$ and $\beta_2$ for a given fermionic calculation.   
%The outcome of this procedure is shown  in Fig. \ref{fig:BLBQBC}a.

In the case of the [4]triangulene dimer, with $s=3/2$, we obtain the following equations:
\begin{eqnarray}
E_\frac{3}{2}(1)-E_\frac{3}{2}(0)= J\left(1-\frac{13}{2}\beta_2 + \frac{511}{16}\beta_3 \right), \\
E_\frac{3}{2}(2)-E_\frac{3}{2}(0)= 3J\left(1-\frac{9}{2}\beta_2 + \frac{279}{16}\beta_3 \right), \\
E_\frac{3}{2}(3)-E_\frac{3}{2}(0)= 6J\left(1-\frac{3}{2}\beta_2 + \frac{171}{16}\beta_3 \right).
\end{eqnarray}
As before, the system of equations can be inverted in order to obtain expressions for $J$, $\beta_2$ and $\beta_3$ in terms of the excitations energies; this then allows us to match the spin model with the fermionic calculation and obtain the dependence of the parameters with $U$ and $t_3$.
%}

%\orangemark{\sout{and the first excited state has $S=1$. In order to assess the open-shell nature of the ground state obtained with CAS calculations, we  employ two methods. First,  compute the weight of the configurations without double occupancy  of the zero modes in the ground state with minimal $S_z$. (See figure X). For $U=0$ the weight is X(Y) for $n=3$($n=4$). As we ramp-up $U$, the weight if find that, for $U>x$ the weight of configurations without double occupancy  increases, reaching ZZ$\%$ for $U=t$.  In our second method we assess the energy separation between the set of lowest energy levels that  can be mapped into a spin model, and the rest of  excited states\cite{Catarina22}. The figure of merit here is the ratio of the gap between the two sets and the energy spread of the lowest energy set. }}

In Fig.~\ref{fig:BLBQBC} we present the values of $J$, $\beta_2$ and $\beta_3$ obtained for the two considered molecules, as a function of $U$, for possible values of $t_3$.
In each panel, the data is only presented up to the critical value of $U$ for which the crossover between low- and high-energy excitations occurs; for smaller $U$ the extraction of the spin model parameters is not valid. 
%the vertical dashed lines mark the critical value of $U$ for which the crossover between low- and high-energy excitations occurs and the energy levels cannot be matched; thus, to the left of this line it is not possible to extract the spin model parameters. 
From these figures, one clearly sees that, for both dimers, the intermolecular antiferromagnetic exchange $J$ is in the order of a few tens of meV, with the $n=3$ dimer presenting a stronger intermolecular exchange than the $n=4$. 
In both molecules we find that for $U \simeq |t|$, the parameter $\beta_2$, which quantifies the weight of the quadratic term relative to the linear one in the model Hamiltonian, takes values up to approximately $1/5$, emphasizing its importance to accurately describe these molecules with a spin model. 
The value of $\beta_3$, describing the strength of the cubic term relative to the leading one, is found to be much smaller than $\beta_2$, indicating that it only introduces a small correction in the spectrum of the [4]triangulene dimer. 
Moreover, the fact that $\beta_3 \ll \beta_2$ further justifies our choice of setting $\beta_3 = 0$ for the $n=3$ dimer.

At last, we note that, for the $n=3$ dimer, the value of $\beta_2$ approaches 1/3 asymptotically as $U$ %goes to zero for all values of $t_3$.
decreases.
In that limit, 
%both the fermion model and 
the spin model---which corresponds to the well-known AKLT model\cite{Affleck1987}---has a vanishing singlet-triplet gap, but it is not a faithful description of the fermion model.
%but the quantum states described by the fermion model are very different and the spin model should not be used to describe the molecules in that limit. 
We note that the singlet-triplet gap cannot vanish for the Hubbard model, as Lieb's theorem\cite{Lieb1989} states that the ground state is unique.

%This is interesting since the BLBQ Hamitlonian with $\beta_2 = 1/3$ corresponds to the well known AKLT model, where the ground state is a quartet. In our case, however, Lieb's theorem guarantees that the ground state of the dimer is always a singlet. which prevents $\beta_2$ from reaching $1/3$ for finite $U$.

\section{Comparison of energy scales controlling open-shell nature of $[n]$triangulene dimers}
\label{app:openshell}

\begin{table}
    \begin{tabular}{ccccc}
$n$ & MO index, $m$ & $\delta$  (meV)        &   IPR & $r$ \\ \hline
3 & 20 (23)  &   199 &   0.139  & 0.53   \\
3 & 21 (22) &   0.2 &     0.140  & $5\times10^{-4}$  \\
4 & 30 (35)  &   197 &    0.132 & 0.55   \\
4 & 31 (34) &   0.2 &    0.092 & $10^{-3}$ \\
4 & 32 (33) &  0    &    0.069 & 0 
\end{tabular}
\caption{
Energy scales for $[n]$triangulene dimers, obtained with $t_3 = 0.1t$. 
MO index $m$ refers to the rank of a given molecular orbital when these are ordered in increasing energy order; the indices in parenthesis refer to the electron-hole symmetric partners.
$\delta$ is the energy splitting between electron-hole symmetric orbitals. 
IPR is the inverse participation ratio defined in Eq.~(\ref{eq: U IPR}) of the sublattice mode of Eq.~(\ref{eq:SLmode}). 
$r$ is defined in Eq.~(\ref{eq:ratio}) and its value, obtained assuming $U=|t|$, indicates the closed- or open-shell nature of a given pair of MOs.
}
\label{tab:ratio}
\end{table}

A preliminary estimate of the open-shell nature of $[n]$triangulene dimers can be obtained by analyzing the ratio between intermolecular hybridization energy of the zero modes and the effective addition energy. 

The low-energy MOs are bonding and antibonding linear combinations of zero modes. 
Therefore, the intermolecular hybridization is proportional to the splitting between electron-hole symmetric single-particle energies, $\pm E_m$, given by $\delta_m=2 |E_m|$. 
The addition energy associated to the double occupancy of a single-triangulene zero mode is given by the product of the atomic Hubbard repulsion parameter $U$ with the inverse participation ratio (IPR)\cite{ortiz19},
\begin{equation}
{\cal U}_m^{(\pm)}=
    U\sum_i |z^{\pm}_m(i)|^4,
    \label{eq: U IPR}
\end{equation}
where the zero mode wave functions can be obtained from the MOs $|m^{\pm}\rangle$, associated to the states with energies $\pm E_m$, through the equation:
\begin{equation}
    |z^{\pm}_m\rangle=\frac{1}{\sqrt{2}}\left(|m^{+}\rangle\pm |m^{-}\rangle\right).
    \label{eq:SLmode}
\end{equation}
%\bluemark{
We note that the zero modes so obtained are not necessarily identical to those obtained from the solution of the individual triangulene problem, on account of the degeneracy of the zero mode manifold, that allows one to define different zero mode bases. 
The values of ${\cal U}$ depend on that choice.
It is found that for centrosymmetric triangulene dimers, ${\cal U}_m^{(+)} = {\cal U}_m^{(-)} \equiv {\cal U}_m$. 
Thus, considering only these two energies, every electron-hole symmetric pair maps into an effective Hubbard model dimer at half-filling, with effective hopping $\tau_m=\delta_m/2$ and Hubbard repulsion ${\cal U}_m$.  
Depending on the ratio between these two energies\cite{malrieu16},
\begin{equation}
 r_m=\frac{\delta_m}{{\cal U}_m}=\frac{2\tau_m}{{\cal U}_m},
 \label{eq:ratio}
\end{equation}
the Hubbard dimer model can describe a closed-shell configuration, for large $r_m$, where two electrons occupy the bonding state with energy $-|E_m|$, or an open-shell system where double occupancy of the zero modes is inhibited. %\bluemark{
We also note that the representation of the many-body Hamiltonian on the zero mode basis contains other interacting terms that couple the effective Hubbard dimers.
%}
%, the low-energy spectrum is identical to a Heisenberg model with $J=4\tau^2/{\cal U}$, and %fractional 
%\purplemark{single } occupancy of both molecular orbitals. \orangemark{This analysis is not really 100\% true for triangulenes. It would be valid for bowties for example. First, in triangulenes linear combinations of two of the almost zero modes orbitals won't give you an actual zero mode back. Second, the picture of purely AF coupled Hubbard dimmers is also not true, because there is FM coupling between them due to the overlap of the wave functions.}

In Table~\ref{tab:ratio} we show the values of $r_m$ for the $n=3$ and $n=4$ triangulene dimers, obtained with $U=|t|$ and $t_3 = 0.1t$. 
All ratios are smaller than 1, in some cases much smaller.  
For comparison, the ratios for the closest energy MOs not formed with zero modes are %$6.5$
$\sim10$ for $n=3$ and %$7$
$\sim11$ for $n=4$. 
This analysis backs up the open-shell nature of triangulene dimers.  

\bibliographystyle{apsrev4-1}
\bibliography{bibshort}

\begin{thebibliography}{61}%
\makeatletter
\providecommand \@ifxundefined [1]{%
 \@ifx{#1\undefined}
}%
\providecommand \@ifnum [1]{%
 \ifnum #1\expandafter \@firstoftwo
 \else \expandafter \@secondoftwo
 \fi
}%
\providecommand \@ifx [1]{%
 \ifx #1\expandafter \@firstoftwo
 \else \expandafter \@secondoftwo
 \fi
}%
\providecommand \natexlab [1]{#1}%
\providecommand \enquote  [1]{``#1''}%
\providecommand \bibnamefont  [1]{#1}%
\providecommand \bibfnamefont [1]{#1}%
\providecommand \citenamefont [1]{#1}%
\providecommand \href@noop [0]{\@secondoftwo}%
\providecommand \href [0]{\begingroup \@sanitize@url \@href}%
\providecommand \@href[1]{\@@startlink{#1}\@@href}%
\providecommand \@@href[1]{\endgroup#1\@@endlink}%
\providecommand \@sanitize@url [0]{\catcode `\\12\catcode `\$12\catcode
  `\&12\catcode `\#12\catcode `\^12\catcode `\_12\catcode `\%12\relax}%
\providecommand \@@startlink[1]{}%
\providecommand \@@endlink[0]{}%
\providecommand \url  [0]{\begingroup\@sanitize@url \@url }%
\providecommand \@url [1]{\endgroup\@href {#1}{\urlprefix }}%
\providecommand \urlprefix  [0]{URL }%
\providecommand \Eprint [0]{\href }%
\providecommand \doibase [0]{http://dx.doi.org/}%
\providecommand \selectlanguage [0]{\@gobble}%
\providecommand \bibinfo  [0]{\@secondoftwo}%
\providecommand \bibfield  [0]{\@secondoftwo}%
\providecommand \translation [1]{[#1]}%
\providecommand \BibitemOpen [0]{}%
\providecommand \bibitemStop [0]{}%
\providecommand \bibitemNoStop [0]{.\EOS\space}%
\providecommand \EOS [0]{\spacefactor3000\relax}%
\providecommand \BibitemShut  [1]{\csname bibitem#1\endcsname}%
\let\auto@bib@innerbib\@empty
%</preamble>
\bibitem [{\citenamefont {Clar}\ and\ \citenamefont {Stewart}(1953)}]{clar53}%
  \BibitemOpen
  \bibfield  {author} {\bibinfo {author} {\bibfnamefont {E.}~\bibnamefont
  {Clar}}\ and\ \bibinfo {author} {\bibfnamefont {D.}~\bibnamefont {Stewart}},\
  }\href@noop {} {\bibfield  {journal} {\bibinfo  {journal} {Journal of the
  American Chemical Society}\ }\textbf {\bibinfo {volume} {75}},\ \bibinfo
  {pages} {2667} (\bibinfo {year} {1953})}\BibitemShut {NoStop}%
\bibitem [{\citenamefont {Fern{\'a}ndez-Rossier}\ and\ \citenamefont
  {Palacios}(2007)}]{fernandez07}%
  \BibitemOpen
  \bibfield  {author} {\bibinfo {author} {\bibfnamefont {J.}~\bibnamefont
  {Fern{\'a}ndez-Rossier}}\ and\ \bibinfo {author} {\bibfnamefont {J.~J.}\
  \bibnamefont {Palacios}},\ }\href@noop {} {\bibfield  {journal} {\bibinfo
  {journal} {Physical Review Letters}\ }\textbf {\bibinfo {volume} {99}},\
  \bibinfo {pages} {177204} (\bibinfo {year} {2007})}\BibitemShut {NoStop}%
\bibitem [{\citenamefont {Su}\ \emph {et~al.}(2020)\citenamefont {Su},
  \citenamefont {Telychko}, \citenamefont {Song},\ and\ \citenamefont
  {Lu}}]{su2020}%
  \BibitemOpen
  \bibfield  {author} {\bibinfo {author} {\bibfnamefont {J.}~\bibnamefont
  {Su}}, \bibinfo {author} {\bibfnamefont {M.}~\bibnamefont {Telychko}},
  \bibinfo {author} {\bibfnamefont {S.}~\bibnamefont {Song}}, \ and\ \bibinfo
  {author} {\bibfnamefont {J.}~\bibnamefont {Lu}},\ }\href@noop {} {\bibfield
  {journal} {\bibinfo  {journal} {Angewandte Chemie International Edition}\
  }\textbf {\bibinfo {volume} {59}},\ \bibinfo {pages} {7658} (\bibinfo {year}
  {2020})}\BibitemShut {NoStop}%
\bibitem [{\citenamefont {Wang}\ \emph {et~al.}(2008)\citenamefont {Wang},
  \citenamefont {Meng},\ and\ \citenamefont {Kaxiras}}]{wang2008}%
  \BibitemOpen
  \bibfield  {author} {\bibinfo {author} {\bibfnamefont {W.~L.}\ \bibnamefont
  {Wang}}, \bibinfo {author} {\bibfnamefont {S.}~\bibnamefont {Meng}}, \ and\
  \bibinfo {author} {\bibfnamefont {E.}~\bibnamefont {Kaxiras}},\ }\href@noop
  {} {\bibfield  {journal} {\bibinfo  {journal} {Nano letters}\ }\textbf
  {\bibinfo {volume} {8}},\ \bibinfo {pages} {241} (\bibinfo {year}
  {2008})}\BibitemShut {NoStop}%
\bibitem [{\citenamefont {Wang}\ \emph {et~al.}(2009)\citenamefont {Wang},
  \citenamefont {Yazyev}, \citenamefont {Meng},\ and\ \citenamefont
  {Kaxiras}}]{wang09}%
  \BibitemOpen
  \bibfield  {author} {\bibinfo {author} {\bibfnamefont {W.~L.}\ \bibnamefont
  {Wang}}, \bibinfo {author} {\bibfnamefont {O.~V.}\ \bibnamefont {Yazyev}},
  \bibinfo {author} {\bibfnamefont {S.}~\bibnamefont {Meng}}, \ and\ \bibinfo
  {author} {\bibfnamefont {E.}~\bibnamefont {Kaxiras}},\ }\href@noop {}
  {\bibfield  {journal} {\bibinfo  {journal} {Physical review letters}\
  }\textbf {\bibinfo {volume} {102}},\ \bibinfo {pages} {157201} (\bibinfo
  {year} {2009})}\BibitemShut {NoStop}%
\bibitem [{\citenamefont {Yazyev}(2010)}]{yazyev10}%
  \BibitemOpen
  \bibfield  {author} {\bibinfo {author} {\bibfnamefont {O.~V.}\ \bibnamefont
  {Yazyev}},\ }\href@noop {} {\bibfield  {journal} {\bibinfo  {journal}
  {Reports on Progress in Physics}\ }\textbf {\bibinfo {volume} {73}},\
  \bibinfo {pages} {056501} (\bibinfo {year} {2010})}\BibitemShut {NoStop}%
\bibitem [{\citenamefont {Ortiz}\ \emph {et~al.}(2019)\citenamefont {Ortiz},
  \citenamefont {Boto}, \citenamefont {Garc{\'\i}a-Mart{\'\i}nez},
  \citenamefont {Sancho-Garc{\'\i}a}, \citenamefont {Melle-Franco},\ and\
  \citenamefont {Fern{\'a}ndez-Rossier}}]{ortiz19}%
  \BibitemOpen
  \bibfield  {author} {\bibinfo {author} {\bibfnamefont {R.}~\bibnamefont
  {Ortiz}}, \bibinfo {author} {\bibfnamefont {R.~{\'A}.}\ \bibnamefont {Boto}},
  \bibinfo {author} {\bibfnamefont {N.}~\bibnamefont
  {Garc{\'\i}a-Mart{\'\i}nez}}, \bibinfo {author} {\bibfnamefont {J.~C.}\
  \bibnamefont {Sancho-Garc{\'\i}a}}, \bibinfo {author} {\bibfnamefont
  {M.}~\bibnamefont {Melle-Franco}}, \ and\ \bibinfo {author} {\bibfnamefont
  {J.}~\bibnamefont {Fern{\'a}ndez-Rossier}},\ }\href@noop {} {\bibfield
  {journal} {\bibinfo  {journal} {Nano Lett.}\ }\textbf {\bibinfo {volume}
  {19}},\ \bibinfo {pages} {5991} (\bibinfo {year} {2019})}\BibitemShut
  {NoStop}%
\bibitem [{\citenamefont {Lieb}(1989)}]{Lieb1989}%
  \BibitemOpen
  \bibfield  {author} {\bibinfo {author} {\bibfnamefont {E.~H.}\ \bibnamefont
  {Lieb}},\ }\href@noop {} {\bibfield  {journal} {\bibinfo  {journal} {Physical
  review letters}\ }\textbf {\bibinfo {volume} {62}},\ \bibinfo {pages} {1201}
  (\bibinfo {year} {1989})}\BibitemShut {NoStop}%
\bibitem [{\citenamefont {Ovchinnikov}(1978)}]{ovchinnikov78}%
  \BibitemOpen
  \bibfield  {author} {\bibinfo {author} {\bibfnamefont {A.~A.}\ \bibnamefont
  {Ovchinnikov}},\ }\href@noop {} {\bibfield  {journal} {\bibinfo  {journal}
  {Theoretica Chimica Acta}\ }\textbf {\bibinfo {volume} {47}},\ \bibinfo
  {pages} {297} (\bibinfo {year} {1978})}\BibitemShut {NoStop}%
\bibitem [{\citenamefont {Cai}\ \emph {et~al.}(2010)\citenamefont {Cai},
  \citenamefont {Ruffieux}, \citenamefont {Jaafar}, \citenamefont {Bieri},
  \citenamefont {Braun}, \citenamefont {Blankenburg}, \citenamefont {Muoth},
  \citenamefont {Seitsonen}, \citenamefont {Saleh}, \citenamefont {Feng} \emph
  {et~al.}}]{cai2010}%
  \BibitemOpen
  \bibfield  {author} {\bibinfo {author} {\bibfnamefont {J.}~\bibnamefont
  {Cai}}, \bibinfo {author} {\bibfnamefont {P.}~\bibnamefont {Ruffieux}},
  \bibinfo {author} {\bibfnamefont {R.}~\bibnamefont {Jaafar}}, \bibinfo
  {author} {\bibfnamefont {M.}~\bibnamefont {Bieri}}, \bibinfo {author}
  {\bibfnamefont {T.}~\bibnamefont {Braun}}, \bibinfo {author} {\bibfnamefont
  {S.}~\bibnamefont {Blankenburg}}, \bibinfo {author} {\bibfnamefont
  {M.}~\bibnamefont {Muoth}}, \bibinfo {author} {\bibfnamefont {A.~P.}\
  \bibnamefont {Seitsonen}}, \bibinfo {author} {\bibfnamefont {M.}~\bibnamefont
  {Saleh}}, \bibinfo {author} {\bibfnamefont {X.}~\bibnamefont {Feng}},  \emph
  {et~al.},\ }\href@noop {} {\bibfield  {journal} {\bibinfo  {journal}
  {Nature}\ }\textbf {\bibinfo {volume} {466}},\ \bibinfo {pages} {470}
  (\bibinfo {year} {2010})}\BibitemShut {NoStop}%
\bibitem [{\citenamefont {Song}\ \emph {et~al.}(2021)\citenamefont {Song},
  \citenamefont {Su}, \citenamefont {Telychko}, \citenamefont {Li},
  \citenamefont {Li}, \citenamefont {Li}, \citenamefont {Su}, \citenamefont
  {Wu},\ and\ \citenamefont {Lu}}]{Song2021}%
  \BibitemOpen
  \bibfield  {author} {\bibinfo {author} {\bibfnamefont {S.}~\bibnamefont
  {Song}}, \bibinfo {author} {\bibfnamefont {J.}~\bibnamefont {Su}}, \bibinfo
  {author} {\bibfnamefont {M.}~\bibnamefont {Telychko}}, \bibinfo {author}
  {\bibfnamefont {J.}~\bibnamefont {Li}}, \bibinfo {author} {\bibfnamefont
  {G.}~\bibnamefont {Li}}, \bibinfo {author} {\bibfnamefont {Y.}~\bibnamefont
  {Li}}, \bibinfo {author} {\bibfnamefont {C.}~\bibnamefont {Su}}, \bibinfo
  {author} {\bibfnamefont {J.}~\bibnamefont {Wu}}, \ and\ \bibinfo {author}
  {\bibfnamefont {J.}~\bibnamefont {Lu}},\ }\href {\doibase 10.1039/D0CS01060J}
  {\bibfield  {journal} {\bibinfo  {journal} {Chem. Soc. Rev.}\ }\textbf
  {\bibinfo {volume} {50}},\ \bibinfo {pages} {3238} (\bibinfo {year}
  {2021})}\BibitemShut {NoStop}%
\bibitem [{\citenamefont {Pavli{\v{c}}ek}\ \emph {et~al.}(2017)\citenamefont
  {Pavli{\v{c}}ek}, \citenamefont {Mistry}, \citenamefont {Majzik},
  \citenamefont {Moll}, \citenamefont {Meyer}, \citenamefont {Fox},\ and\
  \citenamefont {Gross}}]{pavlivcek2017}%
  \BibitemOpen
  \bibfield  {author} {\bibinfo {author} {\bibfnamefont {N.}~\bibnamefont
  {Pavli{\v{c}}ek}}, \bibinfo {author} {\bibfnamefont {A.}~\bibnamefont
  {Mistry}}, \bibinfo {author} {\bibfnamefont {Z.}~\bibnamefont {Majzik}},
  \bibinfo {author} {\bibfnamefont {N.}~\bibnamefont {Moll}}, \bibinfo {author}
  {\bibfnamefont {G.}~\bibnamefont {Meyer}}, \bibinfo {author} {\bibfnamefont
  {D.~J.}\ \bibnamefont {Fox}}, \ and\ \bibinfo {author} {\bibfnamefont
  {L.}~\bibnamefont {Gross}},\ }\href@noop {} {\bibfield  {journal} {\bibinfo
  {journal} {Nature Nanotechnology}\ }\textbf {\bibinfo {volume} {12}},\
  \bibinfo {pages} {308} (\bibinfo {year} {2017})}\BibitemShut {NoStop}%
\bibitem [{\citenamefont {Su}\ \emph {et~al.}(2019)\citenamefont {Su},
  \citenamefont {Telychko}, \citenamefont {Hu}, \citenamefont {Macam},
  \citenamefont {Mutombo}, \citenamefont {Zhang}, \citenamefont {Bao},
  \citenamefont {Cheng}, \citenamefont {Huang}, \citenamefont {Qiu} \emph
  {et~al.}}]{su2019}%
  \BibitemOpen
  \bibfield  {author} {\bibinfo {author} {\bibfnamefont {J.}~\bibnamefont
  {Su}}, \bibinfo {author} {\bibfnamefont {M.}~\bibnamefont {Telychko}},
  \bibinfo {author} {\bibfnamefont {P.}~\bibnamefont {Hu}}, \bibinfo {author}
  {\bibfnamefont {G.}~\bibnamefont {Macam}}, \bibinfo {author} {\bibfnamefont
  {P.}~\bibnamefont {Mutombo}}, \bibinfo {author} {\bibfnamefont
  {H.}~\bibnamefont {Zhang}}, \bibinfo {author} {\bibfnamefont
  {Y.}~\bibnamefont {Bao}}, \bibinfo {author} {\bibfnamefont {F.}~\bibnamefont
  {Cheng}}, \bibinfo {author} {\bibfnamefont {Z.-Q.}\ \bibnamefont {Huang}},
  \bibinfo {author} {\bibfnamefont {Z.}~\bibnamefont {Qiu}},  \emph {et~al.},\
  }\href@noop {} {\bibfield  {journal} {\bibinfo  {journal} {Science advances}\
  }\textbf {\bibinfo {volume} {5}},\ \bibinfo {pages} {eaav7717} (\bibinfo
  {year} {2019})}\BibitemShut {NoStop}%
\bibitem [{\citenamefont {Mishra}\ \emph {et~al.}(2019)\citenamefont {Mishra},
  \citenamefont {Beyer}, \citenamefont {Eimre}, \citenamefont {Liu},
  \citenamefont {Berger}, \citenamefont {Groning}, \citenamefont {Pignedoli},
  \citenamefont {M{\"u}llen}, \citenamefont {Fasel}, \citenamefont {Feng} \emph
  {et~al.}}]{mishra2019b}%
  \BibitemOpen
  \bibfield  {author} {\bibinfo {author} {\bibfnamefont {S.}~\bibnamefont
  {Mishra}}, \bibinfo {author} {\bibfnamefont {D.}~\bibnamefont {Beyer}},
  \bibinfo {author} {\bibfnamefont {K.}~\bibnamefont {Eimre}}, \bibinfo
  {author} {\bibfnamefont {J.}~\bibnamefont {Liu}}, \bibinfo {author}
  {\bibfnamefont {R.}~\bibnamefont {Berger}}, \bibinfo {author} {\bibfnamefont
  {O.}~\bibnamefont {Groning}}, \bibinfo {author} {\bibfnamefont {C.~A.}\
  \bibnamefont {Pignedoli}}, \bibinfo {author} {\bibfnamefont {K.}~\bibnamefont
  {M{\"u}llen}}, \bibinfo {author} {\bibfnamefont {R.}~\bibnamefont {Fasel}},
  \bibinfo {author} {\bibfnamefont {X.}~\bibnamefont {Feng}},  \emph {et~al.},\
  }\href@noop {} {\bibfield  {journal} {\bibinfo  {journal} {Journal of the
  American Chemical Society}\ }\textbf {\bibinfo {volume} {141}},\ \bibinfo
  {pages} {10621} (\bibinfo {year} {2019})}\BibitemShut {NoStop}%
\bibitem [{\citenamefont {Mishra}\ \emph
  {et~al.}(2021{\natexlab{a}})\citenamefont {Mishra}, \citenamefont {Xu},
  \citenamefont {Eimre}, \citenamefont {Komber}, \citenamefont {Ma},
  \citenamefont {Pignedoli}, \citenamefont {Fasel}, \citenamefont {Feng},\ and\
  \citenamefont {Ruffieux}}]{mishra2021b}%
  \BibitemOpen
  \bibfield  {author} {\bibinfo {author} {\bibfnamefont {S.}~\bibnamefont
  {Mishra}}, \bibinfo {author} {\bibfnamefont {K.}~\bibnamefont {Xu}}, \bibinfo
  {author} {\bibfnamefont {K.}~\bibnamefont {Eimre}}, \bibinfo {author}
  {\bibfnamefont {H.}~\bibnamefont {Komber}}, \bibinfo {author} {\bibfnamefont
  {J.}~\bibnamefont {Ma}}, \bibinfo {author} {\bibfnamefont {C.~A.}\
  \bibnamefont {Pignedoli}}, \bibinfo {author} {\bibfnamefont {R.}~\bibnamefont
  {Fasel}}, \bibinfo {author} {\bibfnamefont {X.}~\bibnamefont {Feng}}, \ and\
  \bibinfo {author} {\bibfnamefont {P.}~\bibnamefont {Ruffieux}},\ }\href@noop
  {} {\bibfield  {journal} {\bibinfo  {journal} {Nanoscale}\ }\textbf {\bibinfo
  {volume} {13}},\ \bibinfo {pages} {1624} (\bibinfo {year}
  {2021}{\natexlab{a}})}\BibitemShut {NoStop}%
\bibitem [{\citenamefont {Turco}\ \emph {et~al.}(2023)\citenamefont {Turco},
  \citenamefont {Bernhardt}, \citenamefont {Krane}, \citenamefont {Valenta},
  \citenamefont {Fasel}, \citenamefont {Juri\'{i}c\v{c}ek},\ and\ \citenamefont
  {Ruffieux}}]{turco23}%
  \BibitemOpen
  \bibfield  {author} {\bibinfo {author} {\bibfnamefont {E.}~\bibnamefont
  {Turco}}, \bibinfo {author} {\bibfnamefont {A.}~\bibnamefont {Bernhardt}},
  \bibinfo {author} {\bibfnamefont {N.}~\bibnamefont {Krane}}, \bibinfo
  {author} {\bibfnamefont {L.}~\bibnamefont {Valenta}}, \bibinfo {author}
  {\bibfnamefont {R.}~\bibnamefont {Fasel}}, \bibinfo {author} {\bibfnamefont
  {M.}~\bibnamefont {Juri\'{i}c\v{c}ek}}, \ and\ \bibinfo {author}
  {\bibfnamefont {P.}~\bibnamefont {Ruffieux}},\ }\href@noop {} {\bibfield
  {journal} {\bibinfo  {journal} {JACS Au}\ } (\bibinfo {year}
  {2023})}\BibitemShut {NoStop}%
\bibitem [{\citenamefont {Mishra}\ \emph {et~al.}(2020)\citenamefont {Mishra},
  \citenamefont {Beyer}, \citenamefont {Eimre}, \citenamefont {Ortiz},
  \citenamefont {Fern{\'a}ndez-Rossier}, \citenamefont {Berger}, \citenamefont
  {Gr{\"o}ning}, \citenamefont {Pignedoli}, \citenamefont {Fasel},
  \citenamefont {Feng} \emph {et~al.}}]{mishra2020}%
  \BibitemOpen
  \bibfield  {author} {\bibinfo {author} {\bibfnamefont {S.}~\bibnamefont
  {Mishra}}, \bibinfo {author} {\bibfnamefont {D.}~\bibnamefont {Beyer}},
  \bibinfo {author} {\bibfnamefont {K.}~\bibnamefont {Eimre}}, \bibinfo
  {author} {\bibfnamefont {R.}~\bibnamefont {Ortiz}}, \bibinfo {author}
  {\bibfnamefont {J.}~\bibnamefont {Fern{\'a}ndez-Rossier}}, \bibinfo {author}
  {\bibfnamefont {R.}~\bibnamefont {Berger}}, \bibinfo {author} {\bibfnamefont
  {O.}~\bibnamefont {Gr{\"o}ning}}, \bibinfo {author} {\bibfnamefont {C.~A.}\
  \bibnamefont {Pignedoli}}, \bibinfo {author} {\bibfnamefont {R.}~\bibnamefont
  {Fasel}}, \bibinfo {author} {\bibfnamefont {X.}~\bibnamefont {Feng}},  \emph
  {et~al.},\ }\href@noop {} {\bibfield  {journal} {\bibinfo  {journal}
  {Angewandte Chemie International Edition}\ } (\bibinfo {year}
  {2020})}\BibitemShut {NoStop}%
\bibitem [{\citenamefont {Mishra}\ \emph
  {et~al.}(2021{\natexlab{b}})\citenamefont {Mishra}, \citenamefont {Catarina},
  \citenamefont {Wu}, \citenamefont {Ortiz}, \citenamefont {Jacob},
  \citenamefont {Eimre}, \citenamefont {Ma}, \citenamefont {Pignedoli},
  \citenamefont {Feng}, \citenamefont {Ruffieux}, \citenamefont
  {Fernandez-Rossier},\ and\ \citenamefont {Fasel}}]{Mishra2021}%
  \BibitemOpen
  \bibfield  {author} {\bibinfo {author} {\bibfnamefont {S.}~\bibnamefont
  {Mishra}}, \bibinfo {author} {\bibfnamefont {G.}~\bibnamefont {Catarina}},
  \bibinfo {author} {\bibfnamefont {F.}~\bibnamefont {Wu}}, \bibinfo {author}
  {\bibfnamefont {R.}~\bibnamefont {Ortiz}}, \bibinfo {author} {\bibfnamefont
  {D.}~\bibnamefont {Jacob}}, \bibinfo {author} {\bibfnamefont
  {K.}~\bibnamefont {Eimre}}, \bibinfo {author} {\bibfnamefont
  {J.}~\bibnamefont {Ma}}, \bibinfo {author} {\bibfnamefont {C.~A.}\
  \bibnamefont {Pignedoli}}, \bibinfo {author} {\bibfnamefont {X.}~\bibnamefont
  {Feng}}, \bibinfo {author} {\bibfnamefont {P.}~\bibnamefont {Ruffieux}},
  \bibinfo {author} {\bibfnamefont {J.}~\bibnamefont {Fernandez-Rossier}}, \
  and\ \bibinfo {author} {\bibfnamefont {R.}~\bibnamefont {Fasel}},\
  }\href@noop {} {\bibfield  {journal} {\bibinfo  {journal} {Nature}\ }\textbf
  {\bibinfo {volume} {598}},\ \bibinfo {pages} {287} (\bibinfo {year}
  {2021}{\natexlab{b}})}\BibitemShut {NoStop}%
\bibitem [{\citenamefont {Hieulle}\ \emph {et~al.}(2021)\citenamefont
  {Hieulle}, \citenamefont {Castro}, \citenamefont {Friedrich}, \citenamefont
  {Vegliante}, \citenamefont {Lara}, \citenamefont {Sanz}, \citenamefont {Rey},
  \citenamefont {Corso}, \citenamefont {Frederiksen}, \citenamefont {Pascual}
  \emph {et~al.}}]{hieulle2021}%
  \BibitemOpen
  \bibfield  {author} {\bibinfo {author} {\bibfnamefont {J.}~\bibnamefont
  {Hieulle}}, \bibinfo {author} {\bibfnamefont {S.}~\bibnamefont {Castro}},
  \bibinfo {author} {\bibfnamefont {N.}~\bibnamefont {Friedrich}}, \bibinfo
  {author} {\bibfnamefont {A.}~\bibnamefont {Vegliante}}, \bibinfo {author}
  {\bibfnamefont {F.~R.}\ \bibnamefont {Lara}}, \bibinfo {author}
  {\bibfnamefont {S.}~\bibnamefont {Sanz}}, \bibinfo {author} {\bibfnamefont
  {D.}~\bibnamefont {Rey}}, \bibinfo {author} {\bibfnamefont {M.}~\bibnamefont
  {Corso}}, \bibinfo {author} {\bibfnamefont {T.}~\bibnamefont {Frederiksen}},
  \bibinfo {author} {\bibfnamefont {J.~I.}\ \bibnamefont {Pascual}},  \emph
  {et~al.},\ }\href@noop {} {\bibfield  {journal} {\bibinfo  {journal}
  {Angewandte Chemie International Edition}\ }\textbf {\bibinfo {volume}
  {60}},\ \bibinfo {pages} {25224} (\bibinfo {year} {2021})}\BibitemShut
  {NoStop}%
\bibitem [{\citenamefont {Delgado}\ \emph {et~al.}(2023)\citenamefont
  {Delgado}, \citenamefont {Dusold}, \citenamefont {Jiang}, \citenamefont
  {Cronin}, \citenamefont {Louie},\ and\ \citenamefont {Fischer}}]{delgado23}%
  \BibitemOpen
  \bibfield  {author} {\bibinfo {author} {\bibfnamefont {A.}~\bibnamefont
  {Delgado}}, \bibinfo {author} {\bibfnamefont {C.}~\bibnamefont {Dusold}},
  \bibinfo {author} {\bibfnamefont {J.}~\bibnamefont {Jiang}}, \bibinfo
  {author} {\bibfnamefont {A.}~\bibnamefont {Cronin}}, \bibinfo {author}
  {\bibfnamefont {S.~G.}\ \bibnamefont {Louie}}, \ and\ \bibinfo {author}
  {\bibfnamefont {F.~R.}\ \bibnamefont {Fischer}},\ }\href@noop {} {\bibfield
  {journal} {\bibinfo  {journal} {arXiv preprint arXiv:2301.06171}\ } (\bibinfo
  {year} {2023})}\BibitemShut {NoStop}%
\bibitem [{\citenamefont {Ortiz}\ and\ \citenamefont
  {Fern\'{a}ndez-Rossier}(2020)}]{Ortiz2020}%
  \BibitemOpen
  \bibfield  {author} {\bibinfo {author} {\bibfnamefont {R.}~\bibnamefont
  {Ortiz}}\ and\ \bibinfo {author} {\bibfnamefont {J.}~\bibnamefont
  {Fern\'{a}ndez-Rossier}},\ }\href {\doibase 10.1016/j.progsurf.2020.100595}
  {\bibfield  {journal} {\bibinfo  {journal} {Progress in Surface Science}\
  }\textbf {\bibinfo {volume} {95}},\ \bibinfo {pages} {100595} (\bibinfo
  {year} {2020})}\BibitemShut {NoStop}%
\bibitem [{\citenamefont {Ortiz}\ \emph {et~al.}(2022)\citenamefont {Ortiz},
  \citenamefont {Catarina},\ and\ \citenamefont
  {Fern{\'a}ndez-Rossier}}]{ortiz22}%
  \BibitemOpen
  \bibfield  {author} {\bibinfo {author} {\bibfnamefont {R.}~\bibnamefont
  {Ortiz}}, \bibinfo {author} {\bibfnamefont {G.}~\bibnamefont {Catarina}}, \
  and\ \bibinfo {author} {\bibfnamefont {J.}~\bibnamefont
  {Fern{\'a}ndez-Rossier}},\ }\href@noop {} {\bibfield  {journal} {\bibinfo
  {journal} {2D Materials}\ }\textbf {\bibinfo {volume} {10}},\ \bibinfo
  {pages} {015015} (\bibinfo {year} {2022})}\BibitemShut {NoStop}%
\bibitem [{\citenamefont {Sethi}\ \emph {et~al.}(2021)\citenamefont {Sethi},
  \citenamefont {Zhou}, \citenamefont {Zhu}, \citenamefont {Yang},\ and\
  \citenamefont {Liu}}]{sethi2021}%
  \BibitemOpen
  \bibfield  {author} {\bibinfo {author} {\bibfnamefont {G.}~\bibnamefont
  {Sethi}}, \bibinfo {author} {\bibfnamefont {Y.}~\bibnamefont {Zhou}},
  \bibinfo {author} {\bibfnamefont {L.}~\bibnamefont {Zhu}}, \bibinfo {author}
  {\bibfnamefont {L.}~\bibnamefont {Yang}}, \ and\ \bibinfo {author}
  {\bibfnamefont {F.}~\bibnamefont {Liu}},\ }\href@noop {} {\bibfield
  {journal} {\bibinfo  {journal} {Physical Review Letters}\ }\textbf {\bibinfo
  {volume} {126}},\ \bibinfo {pages} {196403} (\bibinfo {year}
  {2021})}\BibitemShut {NoStop}%
\bibitem [{\citenamefont {Zhou}\ and\ \citenamefont {Liu}(2020)}]{zhou20}%
  \BibitemOpen
  \bibfield  {author} {\bibinfo {author} {\bibfnamefont {Y.}~\bibnamefont
  {Zhou}}\ and\ \bibinfo {author} {\bibfnamefont {F.}~\bibnamefont {Liu}},\
  }\href@noop {} {\bibfield  {journal} {\bibinfo  {journal} {Nano Letters}\
  }\textbf {\bibinfo {volume} {21}},\ \bibinfo {pages} {230} (\bibinfo {year}
  {2020})}\BibitemShut {NoStop}%
\bibitem [{\citenamefont {Fujita}\ \emph {et~al.}(1996)\citenamefont {Fujita},
  \citenamefont {Wakabayashi}, \citenamefont {Nakada},\ and\ \citenamefont
  {Kusakabe}}]{fujita96}%
  \BibitemOpen
  \bibfield  {author} {\bibinfo {author} {\bibfnamefont {M.}~\bibnamefont
  {Fujita}}, \bibinfo {author} {\bibfnamefont {K.}~\bibnamefont {Wakabayashi}},
  \bibinfo {author} {\bibfnamefont {K.}~\bibnamefont {Nakada}}, \ and\ \bibinfo
  {author} {\bibfnamefont {K.}~\bibnamefont {Kusakabe}},\ }\href@noop {}
  {\bibfield  {journal} {\bibinfo  {journal} {Journal of the Physical Society
  of Japan}\ }\textbf {\bibinfo {volume} {65}},\ \bibinfo {pages} {1920}
  (\bibinfo {year} {1996})}\BibitemShut {NoStop}%
\bibitem [{\citenamefont {Wakabayashi}\ \emph {et~al.}(1998)\citenamefont
  {Wakabayashi}, \citenamefont {Sigrist},\ and\ \citenamefont
  {Fujita}}]{wakabayashi98}%
  \BibitemOpen
  \bibfield  {author} {\bibinfo {author} {\bibfnamefont {K.}~\bibnamefont
  {Wakabayashi}}, \bibinfo {author} {\bibfnamefont {M.}~\bibnamefont
  {Sigrist}}, \ and\ \bibinfo {author} {\bibfnamefont {M.}~\bibnamefont
  {Fujita}},\ }\href@noop {} {\bibfield  {journal} {\bibinfo  {journal}
  {Journal of the Physical Society of Japan}\ }\textbf {\bibinfo {volume}
  {67}},\ \bibinfo {pages} {2089} (\bibinfo {year} {1998})}\BibitemShut
  {NoStop}%
\bibitem [{\citenamefont {Peres}\ \emph {et~al.}(2004)\citenamefont {Peres},
  \citenamefont {Ara{\'u}jo},\ and\ \citenamefont {Bozi}}]{peres2004}%
  \BibitemOpen
  \bibfield  {author} {\bibinfo {author} {\bibfnamefont {N.}~\bibnamefont
  {Peres}}, \bibinfo {author} {\bibfnamefont {M.}~\bibnamefont {Ara{\'u}jo}}, \
  and\ \bibinfo {author} {\bibfnamefont {D.}~\bibnamefont {Bozi}},\ }\href@noop
  {} {\bibfield  {journal} {\bibinfo  {journal} {Physical Review B}\ }\textbf
  {\bibinfo {volume} {70}},\ \bibinfo {pages} {195122} (\bibinfo {year}
  {2004})}\BibitemShut {NoStop}%
\bibitem [{\citenamefont {Fern{\'a}ndez-Rossier}(2008)}]{fernandez08}%
  \BibitemOpen
  \bibfield  {author} {\bibinfo {author} {\bibfnamefont {J.}~\bibnamefont
  {Fern{\'a}ndez-Rossier}},\ }\href@noop {} {\bibfield  {journal} {\bibinfo
  {journal} {Physical Review B}\ }\textbf {\bibinfo {volume} {77}},\ \bibinfo
  {pages} {075430} (\bibinfo {year} {2008})}\BibitemShut {NoStop}%
\bibitem [{\citenamefont {Arovas}\ \emph {et~al.}(2022)\citenamefont {Arovas},
  \citenamefont {Berg}, \citenamefont {Kivelson},\ and\ \citenamefont
  {Raghu}}]{Arovas2022}%
  \BibitemOpen
  \bibfield  {author} {\bibinfo {author} {\bibfnamefont {D.~P.}\ \bibnamefont
  {Arovas}}, \bibinfo {author} {\bibfnamefont {E.}~\bibnamefont {Berg}},
  \bibinfo {author} {\bibfnamefont {S.~A.}\ \bibnamefont {Kivelson}}, \ and\
  \bibinfo {author} {\bibfnamefont {S.}~\bibnamefont {Raghu}},\ }\href
  {\doibase 10.1146/annurev-conmatphys-031620-102024} {\bibfield  {journal}
  {\bibinfo  {journal} {Annu. Rev. Condens. Matter Phys.}\ }\textbf {\bibinfo
  {volume} {13}},\ \bibinfo {pages} {239} (\bibinfo {year} {2022})}\BibitemShut
  {NoStop}%
\bibitem [{\citenamefont {Jacob}\ \emph {et~al.}(2021)\citenamefont {Jacob},
  \citenamefont {Ortiz},\ and\ \citenamefont
  {Fern{\'a}ndez-Rossier}}]{jacob21}%
  \BibitemOpen
  \bibfield  {author} {\bibinfo {author} {\bibfnamefont {D.}~\bibnamefont
  {Jacob}}, \bibinfo {author} {\bibfnamefont {R.}~\bibnamefont {Ortiz}}, \ and\
  \bibinfo {author} {\bibfnamefont {J.}~\bibnamefont {Fern{\'a}ndez-Rossier}},\
  }\href@noop {} {\bibfield  {journal} {\bibinfo  {journal} {Physical Review
  B}\ }\textbf {\bibinfo {volume} {104}},\ \bibinfo {pages} {075404} (\bibinfo
  {year} {2021})}\BibitemShut {NoStop}%
\bibitem [{\citenamefont {Jacob}\ and\ \citenamefont
  {Fern{\'a}ndez-Rossier}(2022)}]{jacob22}%
  \BibitemOpen
  \bibfield  {author} {\bibinfo {author} {\bibfnamefont {D.}~\bibnamefont
  {Jacob}}\ and\ \bibinfo {author} {\bibfnamefont {J.}~\bibnamefont
  {Fern{\'a}ndez-Rossier}},\ }\href@noop {} {\bibfield  {journal} {\bibinfo
  {journal} {Physical Review B}\ }\textbf {\bibinfo {volume} {106}},\ \bibinfo
  {pages} {205405} (\bibinfo {year} {2022})}\BibitemShut {NoStop}%
\bibitem [{\citenamefont {Palacios}\ \emph {et~al.}(2008)\citenamefont
  {Palacios}, \citenamefont {Fern{\'a}ndez-Rossier},\ and\ \citenamefont
  {Brey}}]{palacios08}%
  \BibitemOpen
  \bibfield  {author} {\bibinfo {author} {\bibfnamefont {J.~J.}\ \bibnamefont
  {Palacios}}, \bibinfo {author} {\bibfnamefont {J.}~\bibnamefont
  {Fern{\'a}ndez-Rossier}}, \ and\ \bibinfo {author} {\bibfnamefont
  {L.}~\bibnamefont {Brey}},\ }\href@noop {} {\bibfield  {journal} {\bibinfo
  {journal} {Physical Review B}\ }\textbf {\bibinfo {volume} {77}},\ \bibinfo
  {pages} {195428} (\bibinfo {year} {2008})}\BibitemShut {NoStop}%
\bibitem [{\citenamefont {Yazyev}(2008)}]{yazyev08b}%
  \BibitemOpen
  \bibfield  {author} {\bibinfo {author} {\bibfnamefont {O.~V.}\ \bibnamefont
  {Yazyev}},\ }\href@noop {} {\bibfield  {journal} {\bibinfo  {journal}
  {Physical review letters}\ }\textbf {\bibinfo {volume} {101}},\ \bibinfo
  {pages} {037203} (\bibinfo {year} {2008})}\BibitemShut {NoStop}%
\bibitem [{\citenamefont {Jung}\ and\ \citenamefont
  {MacDonald}(2009)}]{jung09}%
  \BibitemOpen
  \bibfield  {author} {\bibinfo {author} {\bibfnamefont {J.}~\bibnamefont
  {Jung}}\ and\ \bibinfo {author} {\bibfnamefont {A.}~\bibnamefont
  {MacDonald}},\ }\href@noop {} {\bibfield  {journal} {\bibinfo  {journal}
  {Physical Review B}\ }\textbf {\bibinfo {volume} {79}},\ \bibinfo {pages}
  {235433} (\bibinfo {year} {2009})}\BibitemShut {NoStop}%
\bibitem [{\citenamefont {Feldner}\ \emph {et~al.}(2010)\citenamefont
  {Feldner}, \citenamefont {Meng}, \citenamefont {Honecker}, \citenamefont
  {Cabra}, \citenamefont {Wessel},\ and\ \citenamefont {Assaad}}]{feldner10}%
  \BibitemOpen
  \bibfield  {author} {\bibinfo {author} {\bibfnamefont {H.}~\bibnamefont
  {Feldner}}, \bibinfo {author} {\bibfnamefont {Z.~Y.}\ \bibnamefont {Meng}},
  \bibinfo {author} {\bibfnamefont {A.}~\bibnamefont {Honecker}}, \bibinfo
  {author} {\bibfnamefont {D.}~\bibnamefont {Cabra}}, \bibinfo {author}
  {\bibfnamefont {S.}~\bibnamefont {Wessel}}, \ and\ \bibinfo {author}
  {\bibfnamefont {F.~F.}\ \bibnamefont {Assaad}},\ }\href@noop {} {\bibfield
  {journal} {\bibinfo  {journal} {Physical Review B}\ }\textbf {\bibinfo
  {volume} {81}},\ \bibinfo {pages} {115416} (\bibinfo {year}
  {2010})}\BibitemShut {NoStop}%
\bibitem [{\citenamefont {Soriano}\ and\ \citenamefont
  {Fern{\'a}ndez-Rossier}(2012)}]{soriano12}%
  \BibitemOpen
  \bibfield  {author} {\bibinfo {author} {\bibfnamefont {D.}~\bibnamefont
  {Soriano}}\ and\ \bibinfo {author} {\bibfnamefont {J.}~\bibnamefont
  {Fern{\'a}ndez-Rossier}},\ }\href@noop {} {\bibfield  {journal} {\bibinfo
  {journal} {Physical Review B}\ }\textbf {\bibinfo {volume} {85}},\ \bibinfo
  {pages} {195433} (\bibinfo {year} {2012})}\BibitemShut {NoStop}%
\bibitem [{\citenamefont {Ij{\"a}s}\ \emph {et~al.}(2013)\citenamefont
  {Ij{\"a}s}, \citenamefont {Ervasti}, \citenamefont {Uppstu}, \citenamefont
  {Liljeroth}, \citenamefont {Van Der~Lit}, \citenamefont {Swart},\ and\
  \citenamefont {Harju}}]{ijas13}%
  \BibitemOpen
  \bibfield  {author} {\bibinfo {author} {\bibfnamefont {M.}~\bibnamefont
  {Ij{\"a}s}}, \bibinfo {author} {\bibfnamefont {M.}~\bibnamefont {Ervasti}},
  \bibinfo {author} {\bibfnamefont {A.}~\bibnamefont {Uppstu}}, \bibinfo
  {author} {\bibfnamefont {P.}~\bibnamefont {Liljeroth}}, \bibinfo {author}
  {\bibfnamefont {J.}~\bibnamefont {Van Der~Lit}}, \bibinfo {author}
  {\bibfnamefont {I.}~\bibnamefont {Swart}}, \ and\ \bibinfo {author}
  {\bibfnamefont {A.}~\bibnamefont {Harju}},\ }\href@noop {} {\bibfield
  {journal} {\bibinfo  {journal} {Physical Review B}\ }\textbf {\bibinfo
  {volume} {88}},\ \bibinfo {pages} {075429} (\bibinfo {year}
  {2013})}\BibitemShut {NoStop}%
\bibitem [{\citenamefont {Ortiz}\ \emph {et~al.}(2018)\citenamefont {Ortiz},
  \citenamefont {Garc\'{i}a-Mart\'{i}nez}, \citenamefont {Lado},\ and\
  \citenamefont {Fern\'{a}ndez-Rossier}}]{Ortiz18}%
  \BibitemOpen
  \bibfield  {author} {\bibinfo {author} {\bibfnamefont {R.}~\bibnamefont
  {Ortiz}}, \bibinfo {author} {\bibfnamefont {N.~A.}\ \bibnamefont
  {Garc\'{i}a-Mart\'{i}nez}}, \bibinfo {author} {\bibfnamefont {J.~L.}\
  \bibnamefont {Lado}}, \ and\ \bibinfo {author} {\bibfnamefont
  {J.}~\bibnamefont {Fern\'{a}ndez-Rossier}},\ }\href {\doibase
  10.1103/PhysRevB.97.195425} {\bibfield  {journal} {\bibinfo  {journal} {Phys.
  Rev. B}\ }\textbf {\bibinfo {volume} {97}},\ \bibinfo {pages} {195425}
  (\bibinfo {year} {2018})}\BibitemShut {NoStop}%
\bibitem [{\citenamefont {Zheng}\ \emph {et~al.}(2020)\citenamefont {Zheng},
  \citenamefont {Li}, \citenamefont {Xu}, \citenamefont {Beyer}, \citenamefont
  {Yue}, \citenamefont {Zhao}, \citenamefont {Wang}, \citenamefont {Guan},
  \citenamefont {Li}, \citenamefont {Zheng} \emph {et~al.}}]{zheng20}%
  \BibitemOpen
  \bibfield  {author} {\bibinfo {author} {\bibfnamefont {Y.}~\bibnamefont
  {Zheng}}, \bibinfo {author} {\bibfnamefont {C.}~\bibnamefont {Li}}, \bibinfo
  {author} {\bibfnamefont {C.}~\bibnamefont {Xu}}, \bibinfo {author}
  {\bibfnamefont {D.}~\bibnamefont {Beyer}}, \bibinfo {author} {\bibfnamefont
  {X.}~\bibnamefont {Yue}}, \bibinfo {author} {\bibfnamefont {Y.}~\bibnamefont
  {Zhao}}, \bibinfo {author} {\bibfnamefont {G.}~\bibnamefont {Wang}}, \bibinfo
  {author} {\bibfnamefont {D.}~\bibnamefont {Guan}}, \bibinfo {author}
  {\bibfnamefont {Y.}~\bibnamefont {Li}}, \bibinfo {author} {\bibfnamefont
  {H.}~\bibnamefont {Zheng}},  \emph {et~al.},\ }\href@noop {} {\bibfield
  {journal} {\bibinfo  {journal} {Nature Communications}\ }\textbf {\bibinfo
  {volume} {11}},\ \bibinfo {pages} {6076} (\bibinfo {year}
  {2020})}\BibitemShut {NoStop}%
\bibitem [{\citenamefont {Son}\ \emph {et~al.}(2006)\citenamefont {Son},
  \citenamefont {Cohen},\ and\ \citenamefont {Louie}}]{son06}%
  \BibitemOpen
  \bibfield  {author} {\bibinfo {author} {\bibfnamefont {Y.-W.}\ \bibnamefont
  {Son}}, \bibinfo {author} {\bibfnamefont {M.~L.}\ \bibnamefont {Cohen}}, \
  and\ \bibinfo {author} {\bibfnamefont {S.~G.}\ \bibnamefont {Louie}},\
  }\href@noop {} {\bibfield  {journal} {\bibinfo  {journal} {Nature}\ }\textbf
  {\bibinfo {volume} {444}},\ \bibinfo {pages} {347} (\bibinfo {year}
  {2006})}\BibitemShut {NoStop}%
\bibitem [{\citenamefont {Barbosa}\ \emph {et~al.}(2001)\citenamefont
  {Barbosa}, \citenamefont {Muniz}, \citenamefont {Costa},\ and\ \citenamefont
  {Mathon}}]{Barbosa2001}%
  \BibitemOpen
  \bibfield  {author} {\bibinfo {author} {\bibfnamefont {L.~H.~M.}\
  \bibnamefont {Barbosa}}, \bibinfo {author} {\bibfnamefont {R.~B.}\
  \bibnamefont {Muniz}}, \bibinfo {author} {\bibfnamefont {A.~T.}\ \bibnamefont
  {Costa}}, \ and\ \bibinfo {author} {\bibfnamefont {J.}~\bibnamefont
  {Mathon}},\ }\href {\doibase 10.1103/PhysRevB.63.174401} {\bibfield
  {journal} {\bibinfo  {journal} {Phys. Rev. B}\ }\textbf {\bibinfo {volume}
  {63}},\ \bibinfo {pages} {174401} (\bibinfo {year} {2001})}\BibitemShut
  {NoStop}%
\bibitem [{\citenamefont {Culchac}\ \emph {et~al.}(2011)\citenamefont
  {Culchac}, \citenamefont {Latgé},\ and\ \citenamefont
  {Costa}}]{Culchac2011}%
  \BibitemOpen
  \bibfield  {author} {\bibinfo {author} {\bibfnamefont {F.~J.}\ \bibnamefont
  {Culchac}}, \bibinfo {author} {\bibfnamefont {A.}~\bibnamefont {Latgé}}, \
  and\ \bibinfo {author} {\bibfnamefont {A.~T.}\ \bibnamefont {Costa}},\ }\href
  {\doibase 10.1088/1367-2630/13/3/033028} {\bibfield  {journal} {\bibinfo
  {journal} {New Journal of Physics}\ }\textbf {\bibinfo {volume} {13}},\
  \bibinfo {pages} {033028} (\bibinfo {year} {2011})}\BibitemShut {NoStop}%
\bibitem [{\citenamefont {Moriya}(1985)}]{MoriyaBook1985}%
  \BibitemOpen
  \bibfield  {author} {\bibinfo {author} {\bibfnamefont {T.}~\bibnamefont
  {Moriya}},\ }\href@noop {} {\emph {\bibinfo {title} {Spin Fluctuations in
  Itinerant Electron Magnetism}}},\ \bibinfo {edition} {1st}\ ed.,\ \bibinfo
  {series} {Springer Series in Solid-State Sciences}, Vol.~\bibinfo {volume}
  {56}\ (\bibinfo  {publisher} {Springer-Verlag},\ \bibinfo {year}
  {1985})\BibitemShut {NoStop}%
\bibitem [{\citenamefont {Giannozzi}\ \emph {et~al.}(2009)\citenamefont
  {Giannozzi}, \citenamefont {Baroni}, \citenamefont {Bonini}, \citenamefont
  {Calandra}, \citenamefont {Car}, \citenamefont {Cavazzoni}, \citenamefont
  {Ceresoli}, \citenamefont {Chiarotti}, \citenamefont {Cococcioni},
  \citenamefont {Dabo}, \citenamefont {Corso}, \citenamefont {de~Gironcoli},
  \citenamefont {Fabris}, \citenamefont {Fratesi}, \citenamefont {Gebauer},
  \citenamefont {Gerstmann}, \citenamefont {Gougoussis}, \citenamefont
  {Kokalj}, \citenamefont {Lazzeri}, \citenamefont {Martin-Samos},
  \citenamefont {Marzari}, \citenamefont {Mauri}, \citenamefont {Mazzarello},
  \citenamefont {Paolini}, \citenamefont {Pasquarello}, \citenamefont
  {Paulatto}, \citenamefont {Sbraccia}, \citenamefont {Scandolo}, \citenamefont
  {Sclauzero}, \citenamefont {Seitsonen}, \citenamefont {Smogunov},
  \citenamefont {Umari},\ and\ \citenamefont {Wentzcovitch}}]{Giannozzi_2009}%
  \BibitemOpen
  \bibfield  {author} {\bibinfo {author} {\bibfnamefont {P.}~\bibnamefont
  {Giannozzi}}, \bibinfo {author} {\bibfnamefont {S.}~\bibnamefont {Baroni}},
  \bibinfo {author} {\bibfnamefont {N.}~\bibnamefont {Bonini}}, \bibinfo
  {author} {\bibfnamefont {M.}~\bibnamefont {Calandra}}, \bibinfo {author}
  {\bibfnamefont {R.}~\bibnamefont {Car}}, \bibinfo {author} {\bibfnamefont
  {C.}~\bibnamefont {Cavazzoni}}, \bibinfo {author} {\bibfnamefont
  {D.}~\bibnamefont {Ceresoli}}, \bibinfo {author} {\bibfnamefont {G.~L.}\
  \bibnamefont {Chiarotti}}, \bibinfo {author} {\bibfnamefont {M.}~\bibnamefont
  {Cococcioni}}, \bibinfo {author} {\bibfnamefont {I.}~\bibnamefont {Dabo}},
  \bibinfo {author} {\bibfnamefont {A.~D.}\ \bibnamefont {Corso}}, \bibinfo
  {author} {\bibfnamefont {S.}~\bibnamefont {de~Gironcoli}}, \bibinfo {author}
  {\bibfnamefont {S.}~\bibnamefont {Fabris}}, \bibinfo {author} {\bibfnamefont
  {G.}~\bibnamefont {Fratesi}}, \bibinfo {author} {\bibfnamefont
  {R.}~\bibnamefont {Gebauer}}, \bibinfo {author} {\bibfnamefont
  {U.}~\bibnamefont {Gerstmann}}, \bibinfo {author} {\bibfnamefont
  {C.}~\bibnamefont {Gougoussis}}, \bibinfo {author} {\bibfnamefont
  {A.}~\bibnamefont {Kokalj}}, \bibinfo {author} {\bibfnamefont
  {M.}~\bibnamefont {Lazzeri}}, \bibinfo {author} {\bibfnamefont
  {L.}~\bibnamefont {Martin-Samos}}, \bibinfo {author} {\bibfnamefont
  {N.}~\bibnamefont {Marzari}}, \bibinfo {author} {\bibfnamefont
  {F.}~\bibnamefont {Mauri}}, \bibinfo {author} {\bibfnamefont
  {R.}~\bibnamefont {Mazzarello}}, \bibinfo {author} {\bibfnamefont
  {S.}~\bibnamefont {Paolini}}, \bibinfo {author} {\bibfnamefont
  {A.}~\bibnamefont {Pasquarello}}, \bibinfo {author} {\bibfnamefont
  {L.}~\bibnamefont {Paulatto}}, \bibinfo {author} {\bibfnamefont
  {C.}~\bibnamefont {Sbraccia}}, \bibinfo {author} {\bibfnamefont
  {S.}~\bibnamefont {Scandolo}}, \bibinfo {author} {\bibfnamefont
  {G.}~\bibnamefont {Sclauzero}}, \bibinfo {author} {\bibfnamefont {A.~P.}\
  \bibnamefont {Seitsonen}}, \bibinfo {author} {\bibfnamefont {A.}~\bibnamefont
  {Smogunov}}, \bibinfo {author} {\bibfnamefont {P.}~\bibnamefont {Umari}}, \
  and\ \bibinfo {author} {\bibfnamefont {R.~M.}\ \bibnamefont {Wentzcovitch}},\
  }\href {\doibase 10.1088/0953-8984/21/39/395502} {\bibfield  {journal}
  {\bibinfo  {journal} {Journal of Physics: Condensed Matter}\ }\textbf
  {\bibinfo {volume} {21}},\ \bibinfo {pages} {395502} (\bibinfo {year}
  {2009})}\BibitemShut {NoStop}%
\bibitem [{\citenamefont {Slater}(1930)}]{Slater1930}%
  \BibitemOpen
  \bibfield  {author} {\bibinfo {author} {\bibfnamefont {J.~C.}\ \bibnamefont
  {Slater}},\ }\href {\doibase 10.1103/PhysRev.36.57} {\bibfield  {journal}
  {\bibinfo  {journal} {Phys. Rev.}\ }\textbf {\bibinfo {volume} {36}},\
  \bibinfo {pages} {57} (\bibinfo {year} {1930})}\BibitemShut {NoStop}%
\bibitem [{\citenamefont {Sutherland}(1986)}]{Sutherland1986}%
  \BibitemOpen
  \bibfield  {author} {\bibinfo {author} {\bibfnamefont {B.}~\bibnamefont
  {Sutherland}},\ }\href@noop {} {\bibfield  {journal} {\bibinfo  {journal}
  {Physical Review B}\ }\textbf {\bibinfo {volume} {34}},\ \bibinfo {pages}
  {5208} (\bibinfo {year} {1986})}\BibitemShut {NoStop}%
\bibitem [{\citenamefont {Catarina}\ and\ \citenamefont
  {Fern\'{a}ndez-Rossier}(2022)}]{Catarina2022}%
  \BibitemOpen
  \bibfield  {author} {\bibinfo {author} {\bibfnamefont {G.}~\bibnamefont
  {Catarina}}\ and\ \bibinfo {author} {\bibfnamefont {J.}~\bibnamefont
  {Fern\'{a}ndez-Rossier}},\ }\href {\doibase 10.1103/PhysRevB.105.L081116}
  {\bibfield  {journal} {\bibinfo  {journal} {Phys. Rev. B}\ }\textbf {\bibinfo
  {volume} {105}},\ \bibinfo {pages} {L081116} (\bibinfo {year}
  {2022})}\BibitemShut {NoStop}%
\bibitem [{\citenamefont {Sorella}\ and\ \citenamefont
  {Tosatti}(1992)}]{sorella92}%
  \BibitemOpen
  \bibfield  {author} {\bibinfo {author} {\bibfnamefont {S.}~\bibnamefont
  {Sorella}}\ and\ \bibinfo {author} {\bibfnamefont {E.}~\bibnamefont
  {Tosatti}},\ }\href@noop {} {\bibfield  {journal} {\bibinfo  {journal} {EPL
  (Europhysics Letters)}\ }\textbf {\bibinfo {volume} {19}},\ \bibinfo {pages}
  {699} (\bibinfo {year} {1992})}\BibitemShut {NoStop}%
\bibitem [{\citenamefont {Fisher}(1971)}]{Fisher1971}%
  \BibitemOpen
  \bibfield  {author} {\bibinfo {author} {\bibfnamefont {B.~E.~A.}\
  \bibnamefont {Fisher}},\ }\href {\doibase 10.1088/0022-3719/4/16/034}
  {\bibfield  {journal} {\bibinfo  {journal} {Journal of Physics C: Solid State
  Physics}\ }\textbf {\bibinfo {volume} {4}},\ \bibinfo {pages} {2695}
  (\bibinfo {year} {1971})}\BibitemShut {NoStop}%
\bibitem [{\citenamefont {Holstein}\ and\ \citenamefont
  {Primakoff}(1940)}]{holstein40}%
  \BibitemOpen
  \bibfield  {author} {\bibinfo {author} {\bibfnamefont {T.}~\bibnamefont
  {Holstein}}\ and\ \bibinfo {author} {\bibfnamefont {H.}~\bibnamefont
  {Primakoff}},\ }\href@noop {} {\bibfield  {journal} {\bibinfo  {journal}
  {Physical Review}\ }\textbf {\bibinfo {volume} {58}},\ \bibinfo {pages}
  {1098} (\bibinfo {year} {1940})}\BibitemShut {NoStop}%
\bibitem [{\citenamefont {Auerbach}(1998)}]{auerbach98}%
  \BibitemOpen
  \bibfield  {author} {\bibinfo {author} {\bibfnamefont {A.}~\bibnamefont
  {Auerbach}},\ }\href@noop {} {\emph {\bibinfo {title} {Interacting electrons
  and quantum magnetism}}}\ (\bibinfo  {publisher} {Springer Science \&
  Business Media},\ \bibinfo {year} {1998})\BibitemShut {NoStop}%
\bibitem [{\citenamefont {Delgado}\ and\ \citenamefont
  {Fern\'andez-Rossier}(2011)}]{delgado11}%
  \BibitemOpen
  \bibfield  {author} {\bibinfo {author} {\bibfnamefont {F.}~\bibnamefont
  {Delgado}}\ and\ \bibinfo {author} {\bibfnamefont {J.}~\bibnamefont
  {Fern\'andez-Rossier}},\ }\href {\doibase 10.1103/PhysRevB.84.045439}
  {\bibfield  {journal} {\bibinfo  {journal} {Phys. Rev. B}\ }\textbf {\bibinfo
  {volume} {84}},\ \bibinfo {pages} {045439} (\bibinfo {year}
  {2011})}\BibitemShut {NoStop}%
\bibitem [{\citenamefont {Spinelli}\ \emph {et~al.}(2014)\citenamefont
  {Spinelli}, \citenamefont {Bryant}, \citenamefont {Delgado}, \citenamefont
  {Fern{\'a}ndez-Rossier},\ and\ \citenamefont {Otte}}]{spinelli14}%
  \BibitemOpen
  \bibfield  {author} {\bibinfo {author} {\bibfnamefont {A.}~\bibnamefont
  {Spinelli}}, \bibinfo {author} {\bibfnamefont {B.}~\bibnamefont {Bryant}},
  \bibinfo {author} {\bibfnamefont {F.}~\bibnamefont {Delgado}}, \bibinfo
  {author} {\bibfnamefont {J.}~\bibnamefont {Fern{\'a}ndez-Rossier}}, \ and\
  \bibinfo {author} {\bibfnamefont {A.~F.}\ \bibnamefont {Otte}},\ }\href@noop
  {} {\bibfield  {journal} {\bibinfo  {journal} {Nature materials}\ }\textbf
  {\bibinfo {volume} {13}},\ \bibinfo {pages} {782} (\bibinfo {year}
  {2014})}\BibitemShut {NoStop}%
\bibitem [{\citenamefont {Klein}\ \emph {et~al.}(2018)\citenamefont {Klein},
  \citenamefont {MacNeill}, \citenamefont {Lado}, \citenamefont {Soriano},
  \citenamefont {Navarro-Moratalla}, \citenamefont {Watanabe}, \citenamefont
  {Taniguchi}, \citenamefont {Manni}, \citenamefont {Canfield}, \citenamefont
  {Fernández-Rossier},\ and\ \citenamefont {Jarillo-Herrero}}]{klein18}%
  \BibitemOpen
  \bibfield  {author} {\bibinfo {author} {\bibfnamefont {D.~R.}\ \bibnamefont
  {Klein}}, \bibinfo {author} {\bibfnamefont {D.}~\bibnamefont {MacNeill}},
  \bibinfo {author} {\bibfnamefont {J.~L.}\ \bibnamefont {Lado}}, \bibinfo
  {author} {\bibfnamefont {D.}~\bibnamefont {Soriano}}, \bibinfo {author}
  {\bibfnamefont {E.}~\bibnamefont {Navarro-Moratalla}}, \bibinfo {author}
  {\bibfnamefont {K.}~\bibnamefont {Watanabe}}, \bibinfo {author}
  {\bibfnamefont {T.}~\bibnamefont {Taniguchi}}, \bibinfo {author}
  {\bibfnamefont {S.}~\bibnamefont {Manni}}, \bibinfo {author} {\bibfnamefont
  {P.}~\bibnamefont {Canfield}}, \bibinfo {author} {\bibfnamefont
  {J.}~\bibnamefont {Fernández-Rossier}}, \ and\ \bibinfo {author}
  {\bibfnamefont {P.}~\bibnamefont {Jarillo-Herrero}},\ }\href {\doibase
  10.1126/science.aar3617} {\bibfield  {journal} {\bibinfo  {journal}
  {Science}\ }\textbf {\bibinfo {volume} {360}},\ \bibinfo {pages} {1218}
  (\bibinfo {year} {2018})},\ \Eprint
  {http://arxiv.org/abs/https://www.science.org/doi/pdf/10.1126/science.aar3617}
  {https://www.science.org/doi/pdf/10.1126/science.aar3617} \BibitemShut
  {NoStop}%
\bibitem [{\citenamefont {Affleck}\ \emph {et~al.}(1987)\citenamefont
  {Affleck}, \citenamefont {Kennedy}, \citenamefont {Lieb},\ and\ \citenamefont
  {Tasaki}}]{Affleck1987}%
  \BibitemOpen
  \bibfield  {author} {\bibinfo {author} {\bibfnamefont {I.}~\bibnamefont
  {Affleck}}, \bibinfo {author} {\bibfnamefont {T.}~\bibnamefont {Kennedy}},
  \bibinfo {author} {\bibfnamefont {E.~H.}\ \bibnamefont {Lieb}}, \ and\
  \bibinfo {author} {\bibfnamefont {H.}~\bibnamefont {Tasaki}},\ }\href
  {\doibase 10.1103/PhysRevLett.59.799} {\bibfield  {journal} {\bibinfo
  {journal} {Phys. Rev. Lett.}\ }\textbf {\bibinfo {volume} {59}},\ \bibinfo
  {pages} {799} (\bibinfo {year} {1987})}\BibitemShut {NoStop}%
\bibitem [{\citenamefont {Wei}\ \emph {et~al.}(2011)\citenamefont {Wei},
  \citenamefont {Affleck},\ and\ \citenamefont {Raussendorf}}]{Wei2011}%
  \BibitemOpen
  \bibfield  {author} {\bibinfo {author} {\bibfnamefont {T.-C.}\ \bibnamefont
  {Wei}}, \bibinfo {author} {\bibfnamefont {I.}~\bibnamefont {Affleck}}, \ and\
  \bibinfo {author} {\bibfnamefont {R.}~\bibnamefont {Raussendorf}},\ }\href
  {\doibase 10.1103/PhysRevLett.106.070501} {\bibfield  {journal} {\bibinfo
  {journal} {Phys. Rev. Lett.}\ }\textbf {\bibinfo {volume} {106}},\ \bibinfo
  {pages} {070501} (\bibinfo {year} {2011})}\BibitemShut {NoStop}%
\bibitem [{\citenamefont {Mermin}\ and\ \citenamefont
  {Wagner}(1966)}]{mermin66}%
  \BibitemOpen
  \bibfield  {author} {\bibinfo {author} {\bibfnamefont {N.~D.}\ \bibnamefont
  {Mermin}}\ and\ \bibinfo {author} {\bibfnamefont {H.}~\bibnamefont
  {Wagner}},\ }\href@noop {} {\bibfield  {journal} {\bibinfo  {journal}
  {Physical Review Letters}\ }\textbf {\bibinfo {volume} {17}},\ \bibinfo
  {pages} {1133} (\bibinfo {year} {1966})}\BibitemShut {NoStop}%
\bibitem [{\citenamefont {Yazyev}\ and\ \citenamefont
  {Katsnelson}(2008)}]{yazyev08}%
  \BibitemOpen
  \bibfield  {author} {\bibinfo {author} {\bibfnamefont {O.~V.}\ \bibnamefont
  {Yazyev}}\ and\ \bibinfo {author} {\bibfnamefont {M.}~\bibnamefont
  {Katsnelson}},\ }\href@noop {} {\bibfield  {journal} {\bibinfo  {journal}
  {Physical Review Letters}\ }\textbf {\bibinfo {volume} {100}},\ \bibinfo
  {pages} {047209} (\bibinfo {year} {2008})}\BibitemShut {NoStop}%
\bibitem [{\citenamefont {Ortiz}\ \emph {et~al.}(2023)\citenamefont {Ortiz},
  \citenamefont {Giedke},\ and\ \citenamefont {Frederiksen}}]{ortiz2023a}%
  \BibitemOpen
  \bibfield  {author} {\bibinfo {author} {\bibfnamefont {R.}~\bibnamefont
  {Ortiz}}, \bibinfo {author} {\bibfnamefont {G.}~\bibnamefont {Giedke}}, \
  and\ \bibinfo {author} {\bibfnamefont {T.}~\bibnamefont {Frederiksen}},\
  }\href@noop {} {\bibfield  {journal} {\bibinfo  {journal} {Physical Review
  B}\ }\textbf {\bibinfo {volume} {107}},\ \bibinfo {pages} {L100416} (\bibinfo
  {year} {2023})}\BibitemShut {NoStop}%
\bibitem [{\citenamefont {Ortiz}(2023)}]{ortiz2023b}%
  \BibitemOpen
  \bibfield  {author} {\bibinfo {author} {\bibfnamefont {R.}~\bibnamefont
  {Ortiz}},\ }\href@noop {} {\bibfield  {journal} {\bibinfo  {journal} {arXiv
  preprint arXiv:2306.05346}\ } (\bibinfo {year} {2023})}\BibitemShut {NoStop}%
\bibitem [{\citenamefont {Malrieu}\ and\ \citenamefont
  {Trinquier}(2016)}]{malrieu16}%
  \BibitemOpen
  \bibfield  {author} {\bibinfo {author} {\bibfnamefont {J.-P.}\ \bibnamefont
  {Malrieu}}\ and\ \bibinfo {author} {\bibfnamefont {G.}~\bibnamefont
  {Trinquier}},\ }\href@noop {} {\bibfield  {journal} {\bibinfo  {journal} {The
  Journal of Physical Chemistry A}\ }\textbf {\bibinfo {volume} {120}},\
  \bibinfo {pages} {9564} (\bibinfo {year} {2016})}\BibitemShut {NoStop}%
\end{thebibliography}%
\end{document}